 \documentclass[letterpaper,twocolumn,10pt]{article} 
 \usepackage{usenix}
 \usepackage{url} 
\usepackage{listings}
\usepackage{multicol}
\usepackage{lipsum}
\usepackage{adjustbox}
\usepackage[font=bf,skip=\baselineskip]{caption}
\usepackage{tabularx}
\usepackage{multirow}
 \usepackage{booktabs}
 \usepackage{graphicx}
 \usepackage{hyperref}
\usepackage{subcaption}
 \usepackage[ruled,vlined,linesnumbered] {algorithm2e}

\usepackage{breakurl} 
\usepackage{boldline}
\usepackage[ruled,vlined]{algorithm2e}
 
\usepackage{pifont}
\usepackage{balance}
\usepackage[htt]{hyphenat}

\newcommand{\an}[2]{{\bf[\textcolor{blue}{#1}: \textcolor{red}{#2}]}}
\newcommand{\JK}[1]{{\an{JK}{#1}}}
\newcommand{\CS}[1]{{\an{CS}{#1}}}
\newcommand{\yj}[1]{{\color{cyan} Yanjie: #1}}
\newcommand{\OL}[1]{\textcolor{blue}{#1}} 
\newcommand{\bheading}[1]{{\vspace{4pt}\noindent{\textbf{#1}}}}

\newcounter{note}[section]

\newcommand{\ignore}[1]{}

\newcommand{\Unity}{\textsf{Unity}\xspace}

\newcounter{packednmbr}

\newenvironment{packeditemize}{
\begin{list}{$\bullet$}{
\setlength{\labelwidth}{8pt}
\setlength{\itemsep}{0pt}
\setlength{\leftmargin}{\labelwidth}
\addtolength{\leftmargin}{\labelsep}
\setlength{\parindent}{0pt}
\setlength{\listparindent}{\parindent}
\setlength{\parsep}{0pt}
\setlength{\topsep}{3pt}}}{\end{list}}

\newcommand{\sysname}{\textsc{AutoVR}\xspace}

\usepackage{color}

\newcommand{\greenote}[2]{{\bf[\textcolor[rgb]{0,0.5,0}{#1}: \textcolor[rgb]{0,0.8,1}{#2}]}}
\newcommand{\ZQ}[1]{{\greenote{ZQ:}{#1}}}

\newcolumntype{R}[2]{%
    >{\adjustbox{angle=#1,lap=\width-(#2)}\bgroup}%
    l%
    <{\egroup}%
}

\newcommand*\rot{\multicolumn{1}{R{30}{1em}}}

  \renewcommand{\paragraph}[1]{\vspace{0.05in}\noindent{\bf{#1}.}}
\AtEndPreamble{
  \usepackage{hyperref}
\hypersetup{
    colorlinks = true,
    linkcolor = blue,
    anchorcolor = blue,
    citecolor = blue,
    filecolor = blue,
    urlcolor = blue
}
}
\begin{document}

\title{\sysname: Revealing Privacy Leaks in Virtual Reality Applications via Grey-Box Fuzzing Techniques in GUI and In-Game Events}

\title{\sysname: Revealing Privacy Leaks in Virtual Reality Applications via \\Automated Event Exploration}

\title{\sysname: Automated User Interface Exploration of Virtual Reality Apps with\\ Applications to Privacy Leak Detection}

\title{\sysname: Automated UI Exploration And Privacy Leak Detection In Virtual Reality Apps}
\pagenumbering{gobble}
\title{\sysname: Automated UI Exploration for Privacy-Sensitive Data Triggering in Virtual Reality Apps}

\title{\sysname: Automated UI Exploration for Detecting Sensitive Data Flow Exposures in Virtual Reality Apps}


\author{
{\rm John Y. Kim\;\;\;
Chaoshun Zuo \;\;\;
Yanjie Zhao \;\;\;
Zhiqiang Lin
}\\
The Ohio State University
} 





\everypar{\looseness=-1}

\maketitle
\begin{abstract}
The rise of Virtual Reality (VR) has provided developers with an unprecedented platform for creating games and applications (apps) that require distinct inputs, different from those of conventional devices like smartphones. The Meta Quest VR platform, driven by Meta, has democratized VR app publishing and attracted millions of users worldwide. However, as the number of published apps grows, there is a notable lack of robust headless tools for user interface (UI) exploration and user event testing. To address this need, we present \sysname, an automatic framework for dynamic UI and user event interaction in VR apps built on the \Unity Engine. Unlike conventional Android and GUI testers, \sysname analyzes the app's internal binary to reveal hidden events, resolves generative event dependencies, and utilizes them for comprehensive exploration of VR apps. Using sensitive data exposure as a performance metric, we compare \sysname with Android \textsf{Monkey}, a widely used headless Android GUI stress testing tool.  Our empirical evaluation demonstrates \sysname's superior performance, triggering an order of magnitude of more sensitive data exposures and significantly enhancing the privacy of VR apps.
%
\looseness=-1
\end{abstract}

\ignore{
The advent of Virtual Reality (VR) has offered developers an unprecedented platform for creating applications (apps) that demand a set of inputs distinct from conventional personal devices. The Oculus VR platform, spearheaded by Meta, has driven the democratization of VR app publishing and attracted millions of users worldwide. As the volume of published apps escalates, the availability of robust tools for user interface (UI) exploration and user event testing falls short. To address this gap, we present AUTOVR, an innovative framework designed for dynamic UI and user event interaction in VR apps built on the Unity Engine. In contrast to conventional Android and GUI testers, AUTOVR delves the app's internal binary to expose otherwise hidden events. Notably, it resolves generative event dependencies and harnesses them to provide a thorough exploration of VR apps.  Utilizing privacy data leaks as a performance metric, we contrast AUTOVR with Android Monkey, a prevalent Android GUI stress testing tool. Our empirical evaluation shows AUTOVR's superior performance, manifesting in its capacity to trigger one order of magnitude more private data functions, thereby contributing a significant step towards enhancing the security and privacy of VR apps.
}

\section{Introduction}
\label{sec:intro}

Virtual Reality (VR) is rapidly emerging as a transformative force in the consumer device market. By 2025, the global VR market is projected to reach USD 32.40 billion, with expectations to soar to USD 187.40 billion by 2030~\cite{VirtualR15:online}. This platform offers users a diverse range of interactions, from physical gestures to haptic feedback, delivering immersive experiences that far surpass traditional platforms like smartphones. As the VR audience continues to expand, the need for rigorous testing and quality assurance of VR apps has become increasingly critical. Ensuring user privacy, safety, and security is paramount: not only to safeguard sensitive data but also to provide authentic and trustworthy user experiences. \looseness=-1

Despite this urgency, automated testing for VR apps remains underdeveloped~\cite{rzig2023virtual}. Conventional tools such as Android \textsf{Monkey}~\cite{monkey}, while capable of UI testing, struggle with the complexities of 3D engine-based apps, particularly those built on \Unity~\cite{unity}, which serves as the backbone for most VR games and applications~\cite{mike_2020}. Notably, \Unity is the only 3D engine officially endorsed by Apple for its Vision Pro VR device~\cite{UnityLau99:online}. However, the absence of accessible source code for VR apps creates a significant gap in understanding their internal workings, demanding a reverse engineering approach to the \Unity VR app binaries~\cite{rzig2023virtual}. \looseness=-1

While reverse engineering of \Unity-centric apps is hardly new, with established success in static analysis~\cite{katycode, 279950} as exemplified by tools like \textsf{Il2CppDumper}~\cite{il2cppdumper} in the \Unity space, the arena of dynamic analysis tools specifically designed for \Unity apps remains vacant. \Unity-based apps, such as video games, not only present a complex matrix for security testing due to event dependencies on preceding user actions as well as in-app physical interactions but also contain a broader amount of execution entry points, meaning that there is more than one way to interact with the ``main" execution of the app. 
Therefore, current security testing tools such as UI fuzzing, are often ineffective on \Unity-based apps. \looseness=-1

Upon detailed examination of \Unity app binaries, four distinct observations surface: (1) the \Unity app UI structure is embedded within the \Unity app binary, instead of traditional markup or configuration files; 
(2) the \Unity UI structure is generative: UI objects and in-app objects become enabled or disabled based on the state of the app execution;
(3) the binary's operation on a virtual machine that offers internal APIs, thereby enabling the identification of class, method, and object symbols within the \Unity app at run-time; 
(4) \Unity VR apps are highly reliant on physical interactions (e.g., grabbing, hitting, and moving) with surrounding in-app objects, such interactions may contain critical functionality as users are constantly interacting with such objects. \looseness=-1

In light of these observations, we present \sysname, the first automated UI exploration tool that leverages \Unity's unique generative features and internal binary introspection APIs. \sysname ambitiously tackles three interrelated challenges from \Unity VR apps: (1) the semantic recovery of the \Unity UI information, 
(2) the modeling of generative \Unity UI elements to extract event handlers, and (3) the context-aware execution of these handlers by resolving event dependencies.
As a practical demonstration, we utilize \sysname to detect sensitive data exposures that developers may have embedded within their apps. 


We have developed \sysname on top of \textsf{Frida}~\cite{frida}, an open-source dynamic instrumentation toolkit, compatible with Android devices and, consequently, Meta Quest devices. To gauge the effectiveness of \sysname, we have evaluated its performance against the widely used tool Android \textsf{Monkey} across the \Unity VR apps, evaluating the sensitive data exposures that both tools could invoke. In this empirical assessment spanning 366 apps, including 103 paid apps, \sysname demonstrated a remarkable capacity to activate 390 \textbf{unique} sensitive data exposures, dwarfing the 117 instances (a 2.2$\times$ coverage increase) induced by \textsf{Monkey}. \looseness=-1  


\bheading{Contributions.} We make the following contributions:
\begin{packeditemize}
    \item {\bf Novel Techniques (\S\ref{sec:design}).}  We introduce novel techniques including VR \textit{UI semantic recovery}, \textit{generative event modeling} to recover symbols and semantics of \Unity VR apps, and  \textit{context-aware event execution} of UI elements. 
    \looseness=-1


    \item {\bf Practical Framework  (\S\ref{sec:implementation}).} We have developed these techniques in \sysname, an open-source\footnote{https://github.com/OSUSecLab/AutoVR} framework for automated UI exploration of VR apps,  compatible with the ever-changing environment of VR applications.
    
    \item {\bf Empirical Evaluation  (\S\ref{sec:evaluation}).} Our evaluation with 366 VR apps shows a superior capacity of \sysname to trigger 2.2x more \textbf{unique} sensitive data exposing functions in comparison to widely used tool \textsf{Monkey}. \looseness=-1
\end{packeditemize}

\section{Background}
\label{sec:background}

\ignore{
\begin{figure}
    \centering
    \includegraphics[width=0.49\textwidth]{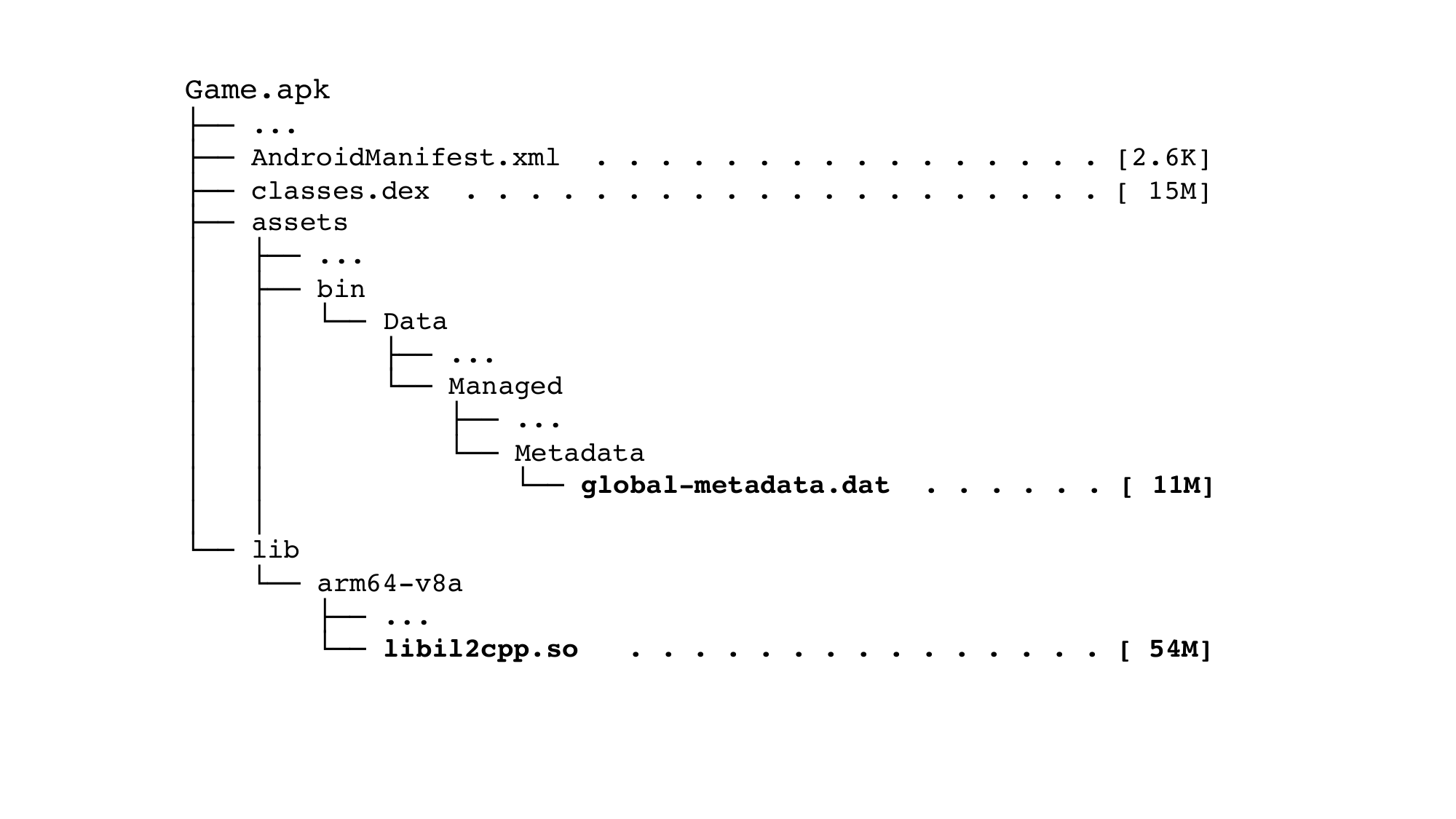}
    \caption{File structure of a Unity APK file, highlighting the game binary and metadata file.}
\label{fig:fs}
\end{figure}
}

Developing VR apps is a complex task, often requiring specialized tools and support. 3D engines like \Unity have made this process much easier by providing essential tools such as real-time rendering APIs, documentation, cross-platform support, and user-friendly IDEs. 
In fact, \Unity is the most widely used game engine within the Meta Quest store~\cite{ovrstore}. Given its popularity and extensive support for VR development, we have chosen to focus on VR apps developed using the \Unity engine.

\paragraph{Scripting Backend} \Unity VR apps, much like mobile games, are developed using C\#, with the final binary translated into assembled C++ code. Developers may choose from two translation scripting backends: (1) \textsf{Mono}, a just-in-time (JIT) compiler~\cite{mono}, or (2) \textsf{IL2CPP}, an ahead-of-time (AOT) compiler~\cite{il2cpp}. \Unity predominantly emphasizes \textsf{IL2CPP} for its enhanced performance and security. 


\paragraph{IL2CPP} The translation of C\# code to C++ is a complex process, requiring compatibility with specific C\# features such as garbage collection, reflection, and exception handling~\cite{billwagner}. Both developer scripts and \Unity libraries written in C\# must be translated to C++. \textsf{IL2CPP} accomplishes this translation, augmenting each function with additional checks. The resultant assembled native C++ binary, \texttt{libil2cpp.so}, has symbols (e.g., class names, method names, and field names) stripped~\cite{stripping}, but to support reflection and other features, the symbols are encrypted and stored separately in \texttt{global-metadata.dat}, bundled within the APK. \looseness=-1

\paragraph{Scene} Within \Unity, a scene acts as a container for game objects, each defined with specific attributes such as position, rotation, and scale. Managed by the \texttt{SceneManager} object~\cite{scenemanager}, scenes are analogous to the UI in Android's \texttt{Activity}. Although an app may include multiple scenes, only one is rendered at a time, allowing \Unity UI to switch between them to create varied app scenarios.


\paragraph{GameObject} \texttt{GameObjects} in \Unity~\cite{gameobject, gameobject1} can represent any object within the game environment, such as players, walls, or UI buttons. Serving as the building blocks of a \Unity app, each \texttt{GameObject} can host multiple \texttt{Component} objects in a modular fashion~\cite{component}. These components govern the \texttt{GameObject}'s behavior, such as movement or sound, and can range in complexity. \Unity provides numerous base components to control app logic, including \texttt{Collider}~\cite{collider} components for collision detection and the \texttt{UnityEngine.UI} components for UI handling~\cite{unityui}.

\begin{figure}[t]
\centering
\subfloat[{A Trigger Event}]{\includegraphics[width=0.2\textwidth]{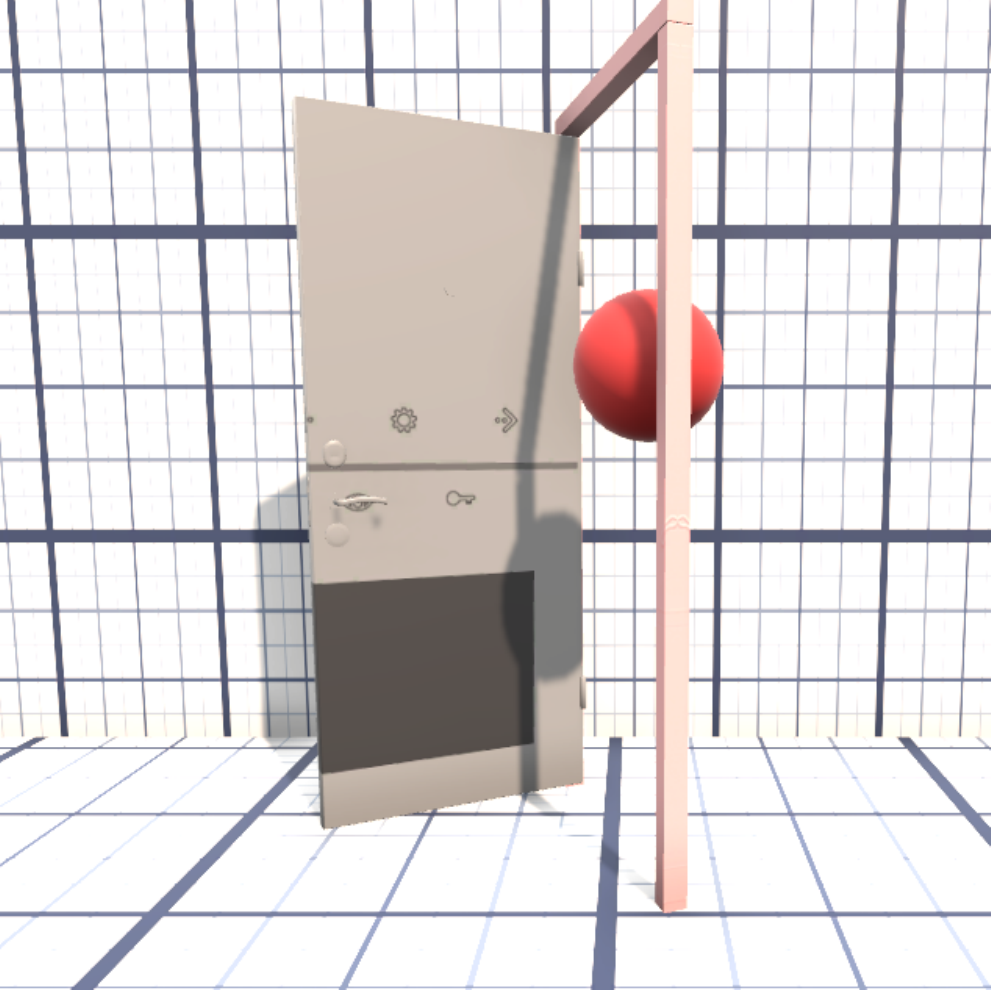}\label{eval:compos}}
\hfill
\subfloat[{A Collision Event}]{\includegraphics[width=0.2\textwidth]{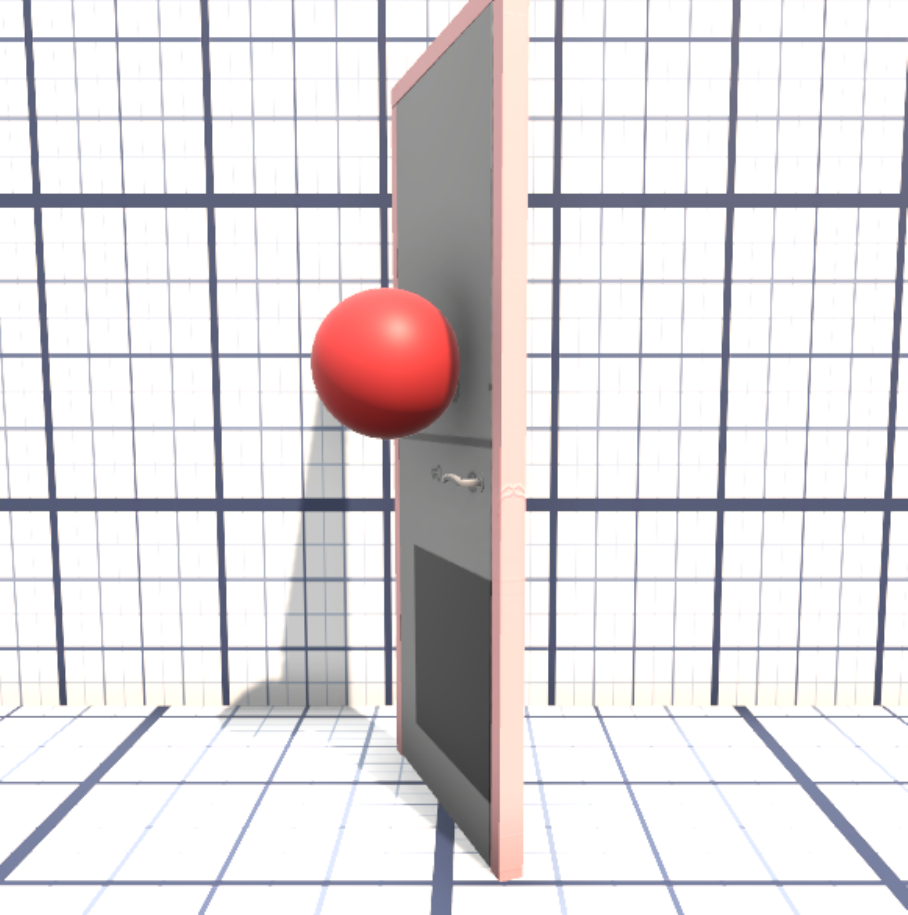}\label{eval:compts}}
 \vspace{-0.15in}
\caption{\textbf{Illustration of Trigger and Collision Events}}
\label{eval:bytecomp}
\end{figure}

\begin{figure*}[h]
   \vspace{-0.2in}
    \centering
        \includegraphics[width=0.7\textwidth]{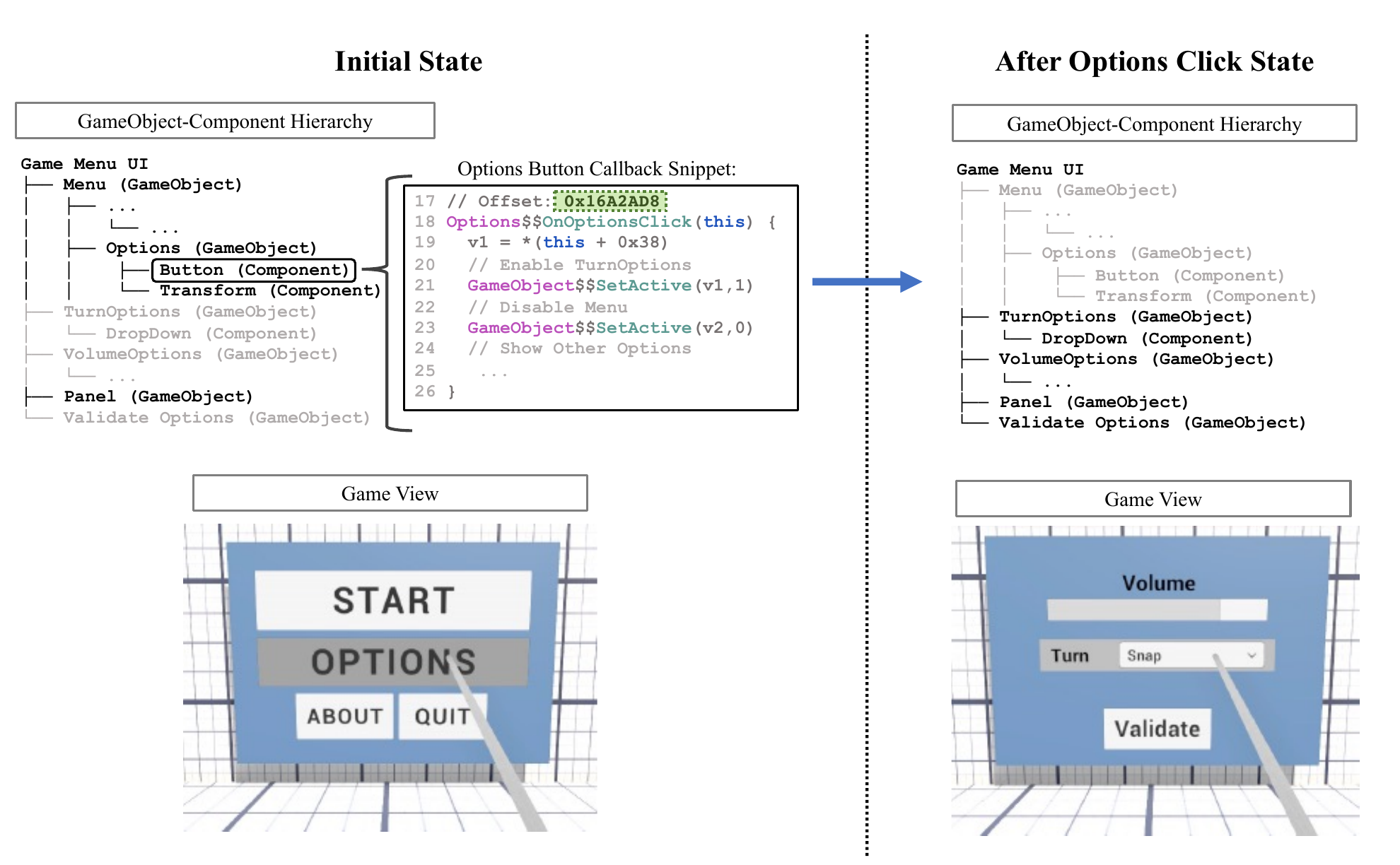}
          \vspace{-0.15in}
    \caption{Running example of a \Unity Scene, illustrating Unity UI logic internals. 
    }
    \vspace{-0.1in}
\label{fig:runex}
\end{figure*}

\paragraph{UI Events}
UI events are typically user-driven and can be triggered through interaction with the VR controllers, such as pointing and clicking using the controller's trigger button. A detailed illustration of UI events can be found in \autoref{fig:runex}. This example presents a series of \texttt{GameObject}s---such as the \texttt{START} button, \texttt{OPTIONS}, \texttt{ABOUT}, and \texttt{QUIT}---each equipped with a \texttt{Button} component that manages their appearance and logic within the VR app. In this example, we assume a developer crafting this user interface in the \Unity Editor by attaching a \texttt{Button} component~\cite{Button1, Button2} to each button-themed \texttt{GameObject}. While other UI elements also exist, buttons are universally understood and functionally unambiguous.  \looseness=-1 

\paragraph{Physics Events}
In addition to UI events, there are physics events that involve interactions with 3D objects within the VR play area and may be associated with critical functionalities, such as scene changes. There are two primary kinds of physics events, as illustrated in~\autoref{eval:bytecomp}:  
\begin{packeditemize}
\item [\textbf{(1) Triggers.}] Triggers are activated when one \texttt{GameObject} intersects with the bounds of another~\cite{events, colliders}. Unlike collisionable objects, trigger objects are not solid, allowing other \texttt{GameObject}s to pass through rather than collide. For instance, a 3D ball object is thrown through an open door. The ball intersects the bounds of the door but is not physically affected by the door. Such an event is classified as a trigger event. \looseness=-1

\item [\textbf{(2) Collisions.}] Collisions happen when a solid \texttt{GameObject} intersects with the bounds of another solid object that is \textbf{not} a trigger~\cite{events, colliders}. For example, a solid enemy hand might interact with a rock, another solid object, by picking it up, touching it, or engaging in other forms of interaction. In contrast to the ball and open door example, suppose the door was closed and is now a solid, non-trigger object. Now the ball physically bounces off the door on intersection, and such an event is classified as a collision event. 
\end{packeditemize}

\section{Overview}
\label{sec:overview}
\vspace{-0.05in}
\subsection{Objective and Scope} 

\vspace{-0.08in}
\noindent\textbf{Objective.} The primary objective of this work is to develop \sysname, an automated UI testing framework tailored specifically for 3D VR apps developed using \Unity. Unlike traditional Android app UI exploration, this framework can effectively test \Unity apps, which are packaged as Android Package files (APK). Traditional methods stumble because UI/physics events within \Unity apps are concealed within the binary, rendering them neither parsable nor adaptable to standard UI exploration techniques. \sysname seeks to overcome these barriers, with an extended ambition to employ the framework for enhancing security and privacy. We showcase the effectiveness of our framework by systematically analyzing privacy data exposure in third-party \Unity VR apps. \looseness=-1


\paragraph{Scope} While there are different 3D engine-based VR apps, this work focuses on VR apps based on the \Unity Engine, specifically those utilizing \textsf{\textsf{IL2CPP}}. As mentioned in \S\ref{sec:background}, the majority of VR apps for Quest devices are developed via the \Unity Engine---a trend likely to escalate with the introduction of new VR devices like the Apple Vision Pro, which employs \Unity for the app development~\cite{UnityLau99:online}. For this work, we focus on \Unity apps compiled using the \textsf{\textsf{IL2CPP}} 
compiler. Compared to the JIT compiler \texttt{Mono}, \textsf{\textsf{IL2CPP}}, an AOT compiler, ensures enhanced performance, a critical attribute for VR apps. Additionally, \textsf{\textsf{IL2CPP}} is anticipated to be used by future \Unity-developed VR apps~\cite{Hauwert_2014}. \looseness=-1


\subsection{Running Example}

To better understand the \Unity UI structure, we depict a VR UI example formulated using the \Unity Editor; we have shown this example in \autoref{fig:runex}.
%
%
The left side of the figure shows the initial state of the UI app before the user clicks the \texttt{OPTIONS} button. The right-hand side shows the state of the app after the \texttt{OPTIONS} button is clicked. 
We also illustrate the hierarchical interconnection of \texttt{GameObject}s and their corresponding components within the app. This hierarchy may encompass child \texttt{GameObject}s. Grayed-out text represents currently deactivated \texttt{GameObject}s, whereas bold text signifies enabled ones. The \textit{Game View} segment reflects what a user would perceive within his/her VR device upon executing this app at their respective state. 
The highlighted portion (i.e., \texttt{0x16A2AD8}) indicates the function offsets or virtual addresses of the UI element's function callback. \looseness=-1

\begin{figure}[t]
\centering
\scriptsize
\graphicspath{{./Figs/}}
\includegraphics[width=.475\textwidth, keepaspectratio]{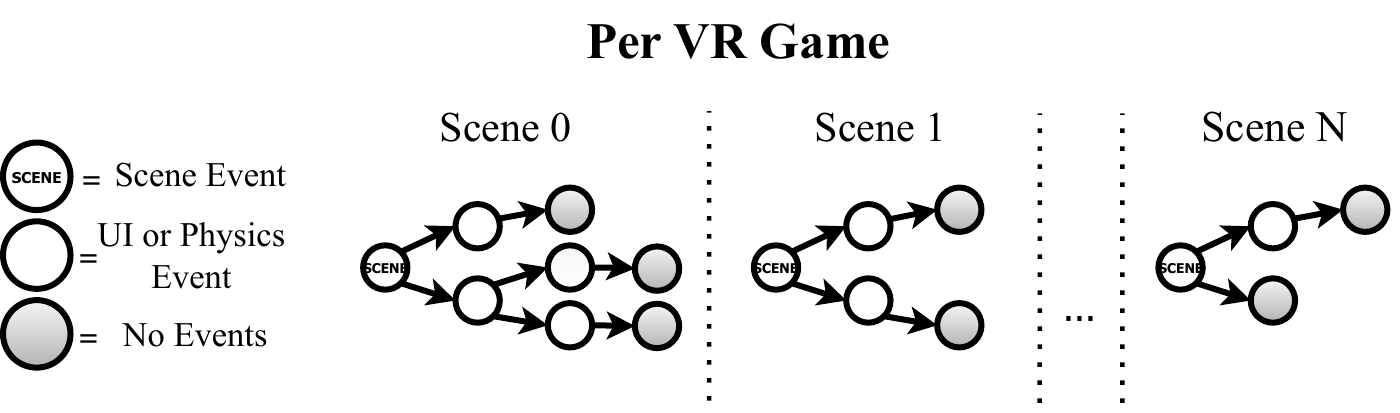}
         \vspace{-0.15in}
\caption{Outline of a scene model per VR game.}
\label{fig:per_vr_game}
 \vspace{-0.15in}
\end{figure}

We also present the code snippet for the \texttt{OPTIONS} button's event function callback (EFC). The \texttt{OPTIONS} button's EFC (denoted by \texttt{Options\$\$OnOptionsClick}) activates the \texttt{TurnOptions} \texttt{GameObject}, along with its associated components, thereby making them visible to the user within the UI. Simultaneously, this action leads to the disabling of the previous \texttt{Menu} \texttt{GameObject}, and their subsequent child \texttt{GameObject}s, as indicated in line 23 of the ``{Options Button Callback Snippet}''.


\subsection{Challenges and Insights}
\label{sec:chall}

\begin{figure*}[h]
    \centering
        \includegraphics[width=0.8\textwidth]{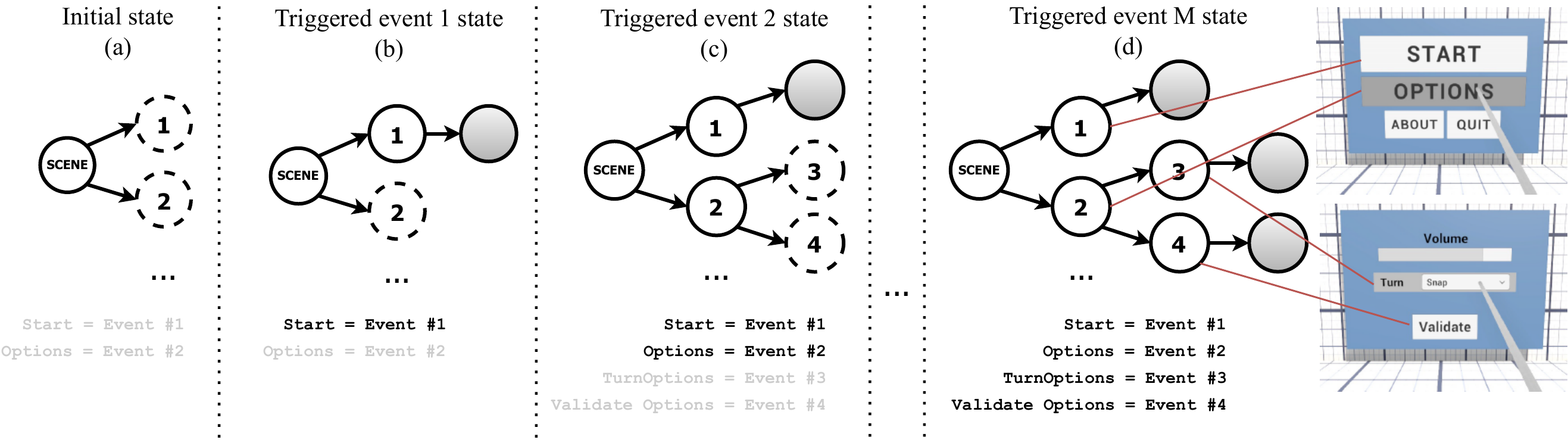}
          \vspace{-0.1in}
    \caption{Generative UI-driven model of the running example. This is analogous to Scene 0 in \autoref{fig:per_vr_game}, where the modeling process per scene is repeated for every scene in a VR game. The lighter colored text indicates the UI event has been found but not yet triggered, whereas the darkened colored text indicates a the UI event has been triggered.}
\label{fig:deps}
\end{figure*}


   \vspace{-0.08in}
\noindent\textbf{(C1) How to Recover UI Semantics.} 
Parsing \Unity UI elements differs from typical Android apps as these elements are not directly accessible through the Android app layer nor from a static configuration file such as the \texttt{AndroidManifest.xml} file. For instance, when using the Android UI Inspector tool in \textsf{Android Studio}, it shows the UI hierarchy of all Android UI elements. However, when debugging a \Unity app, this inspector will not display any UI elements created within the \Unity app engine. This is because \Unity apps do not embed UI elements through Android activities, instead they are embedded within the \textsf{\textsf{IL2CPP}} game binary. \looseness=-1

To solve this challenge, we employ \textsf{\textsf{Frida}} to dynamically instrument the \textsf{IL2CPP} app binary and extract UI elements at runtime. \textsf{IL2CPP} provides a runtime introspection API utilized by the \Unity Engine to access detailed class metadata, including class fields and methods, and to create, collect, and resolve object types at any state of the \Unity app. While these APIs resemble standard C\# libraries, they have been modified by the \textsf{IL2CPP} compiler for optimizations and C++ compatibility. Unlike previous approaches~\cite{katycode, 279950} that statically accessed class metadata from the \texttt{libil2cpp.so} binary using \texttt{global-metadata.dat}, our solution directly invokes \textsf{IL2CPP} API functions from the \texttt{libil2cpp.so} binary using \textsf{Frida} to access class metadata. However, while direct introspection can contain accurate symbols, critical symbols are sometimes lost as some functions are called only through reflection rather than direct invocation. As such, we combine both the static \texttt{global-metadata.dat} file along with dynamic introspection, to get a comprehensive list of \textsf{IL2CPP} functions to call.\looseness=-1



\paragraph{(C2.1) How to Parse and Model UI Events} To extract and model the \Unity UI structure, parsing the \Unity UI is necessary. Unlike typical Android apps, where the UI is separate from the app logic, \Unity UI elements are integrated directly into the logic and perform like generic \texttt{GameObject}s. This relationship is illustrated by the hierarchy in \autoref{fig:runex}, where GUI components like the \texttt{Button} are attached to their respective \texttt{GameObject}s, much like other objects in an app. Therefore, we face the challenge of collecting UI elements and their callbacks from a generative UI at runtime, rather than relying on static configurations. 
Furthermore, the extraction of the function callbacks from the UI elements poses an additional challenge: the \Unity SDK provides two different ways of assigning EFCs to UI event components: dynamically by using the \texttt{UnityEvent.AddListener} API, or static assignment in the \Unity Editor. \looseness=-1

We identify UI elements and event handling interfaces in \Unity by identifying the \texttt{IEventSystemHandler} interface in Component classes using the \textsf{IL2CPP} class introspection API. We bypass the time-consuming process of class inspection by directly accessing loaded objects within the app. This allows to only extract the UI elements in the current scene. We then analyze each object at the start of a Scene to identify event system handler objects as UI elements, extracting EFCs. Simply using API hooks for the \texttt{UnityEvent} is insufficient, as function callbacks assigned in the editor will not use the \texttt{UnityEvent.AddListener} API, nor any API during initialization. As such, to extract the hidden function callbacks, we use \textsf{IL2CPP}'s field introspection API to extract the developer's function callbacks, enabling the extraction of UI objects at runtime. We call this process \textit{Generative UI Modeling}. \looseness=-1

\paragraph{(C2.2) How to Parse and Model Physics Events} 
Similarly, we realize physics events are also generative, especially in a VR environment. For example, we notice a popular game \textit{VRChat} utilizes \textit{physics} events to perform scene changes (e.g., when players enter portals to different online rooms). In a VR 3D environment, it is crucial to account for physics, as a significant portion of VR interactions involve the movement and intersection of objects. 

Fortunately, physics events are comparatively easier to extract than UI events. As outlined in \S\ref{sec:background}, \Unity has two main physics event types: \textit{collision} and \textit{trigger}. Physics events occur when two \texttt{GameObject}s (with an attached Collider component) intersect. To trigger these events, \texttt{GameObject}s must follow their respective physics type's rules, in order to trigger the correct event. As such, to invoke \texttt{Collision} type events, each respective GameObject must contain both a \texttt{Collider} and \texttt{RigidBody} component. To invoke 
\textit{Trigger} type events, each component must have an attached \texttt{Collider} component, and the \texttt{isTrigger} property set to be \textbf{true}. Using \texttt{\textsf{IL2CPP}} and \texttt{\textsf{Frida}}, we can use semantic information resolved from \textbf{C1} to find \texttt{GameObject}s that match their respective physics criteria and set the positions of each \texttt{GameObject} to intersect one another. We iteratively perform these intersections on dynamic \texttt{GameObject}s within the scene. As such, we similarly label this process \textit{Generative Physics Modeling}.

\begin{figure}[tb]
\centering
\scriptsize
\graphicspath{{./Figs/}}
\includegraphics[width=.4\textwidth, keepaspectratio]{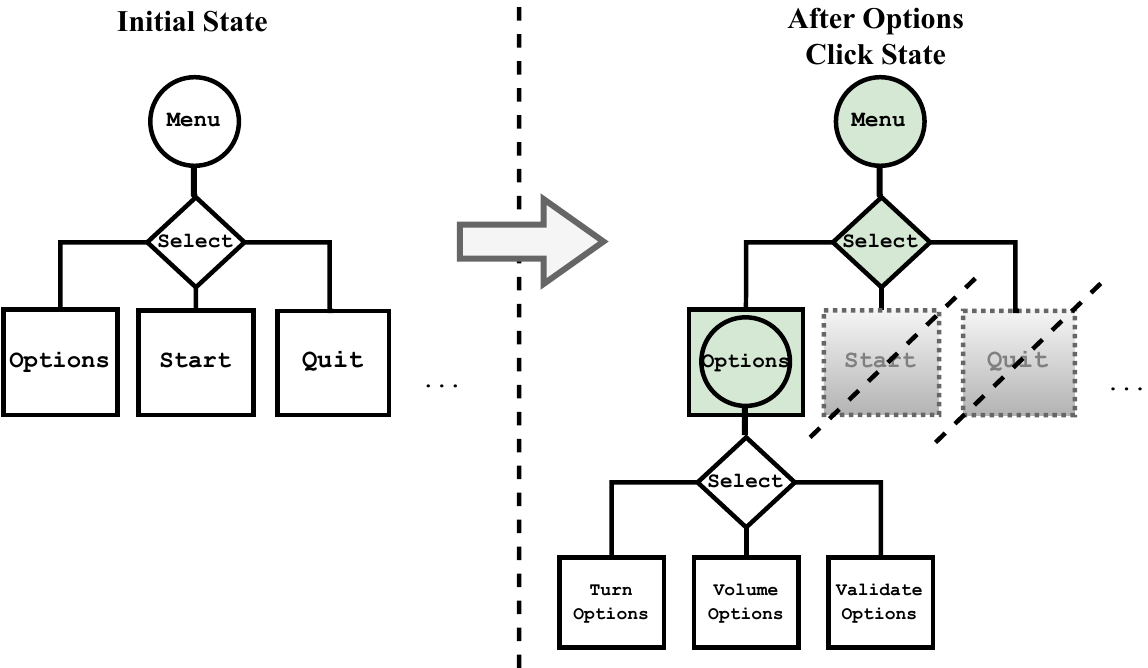}
\vspace{-0.1in}
\caption{High level event state before and after \texttt{Options\$\$OnOptionsClick} was clicked in \autoref{fig:runex}. Highlighted in green is the path taken, and highlighted gray with dotted outlines are disabled events. }
\label{fig:loss}
\vspace{-0.1in}
\end{figure}

\paragraph{(C3) How to Execute Events and Resolve Event Dependencies}
Executing the EFCs from the generative \Unity UI also introduces another challenge: dynamic dependencies. Because the UI is generative, it is also possible that triggering EFCs may cause other UI \texttt{GameObject}s to be deactivated, or spawn/enable new UI \texttt{GameObject}s. For instance, in \autoref{fig:runex}, we notice that the \texttt{Options\$\$OnOptionsClick} will disable/deactivate the Menu UI and all UI elements underneath. However, neither \texttt{About} nor \texttt{Quit} buttons have been explored. We identify these issues as dynamic dependencies where the spawned or deactivated UI \texttt{GameObject}s are dependent on the initial UI \texttt{GameObject} executed. This also applies for \texttt{GameObject}s that integrate physics events. This dependency model occurs for every scene within a VR game, as such, each scene may contain an entirely different set of events to model. \autoref{fig:per_vr_game} illustrates example event models for each scene per VR game, representing the complete event exploration state of the game. To be complete, execution of as-many-as-possible modeled event function callbacks is necessary and is a challenge we solve for this work. 

We solve this challenge by creating a UI-driven generative state model, where the triggering of one such event may cause more unknown events to be spawned or enabled, thus adding to the overall known event state space of a scene. We showcase this process in  \autoref{fig:deps}. When the initial scene is loaded (a), the game uncovers two events that have not been triggered. In a depth-first search manner, each event is triggered, and, as a side effect, uncover new unknown (or disabled) events. Additionally, it is possible that no additional events may be uncovered, leading to an empty set of next-events (b). However, unlike a typical tree-traversal problem, the state of the game cannot be easily back-tracked to its previous state, as the invocation of one event may lose the existence of another event. We illustrate this problem in \autoref{fig:loss}. When the \texttt{OPTIONS} button is selected, the \texttt{START} and \texttt{QUIT} buttons become disabled and cannot be further exercised upon, losing valuable events. As such, to solve this, we utilize a synthetic Scene Event that acts as the state reset in the game. This scene event essentially loads/reloads the current scene by leveraging \texttt{Frida}'s unique capabilities of function invocation on the \textsf{IL2CPP} binary. Specifically, we notice that \Unity's core scene invocation relies on the \texttt{LoadSceneAsyncNameIndexInternal} function to load new scenes. Upon the invocation of this function, it is possible to reload the scene back to its initial state, effectively ``back-tracking" the scene state to explore a different event path. Using the information gathered by (a) and (b) in \autoref{fig:deps}, we can derive the next sequence of unexplored events to trigger, without losing the information of other events, leading to state (c). Eventually, this process is repeated until every event path leads to an empty set of new events, ensuring the maximum number events are exercised per scene.

\section{Detailed Design}
\label{sec:design}

\begin{figure}[t]
\centering
\scriptsize
\graphicspath{{./Figs/}}
\includegraphics[width=.485\textwidth, keepaspectratio]{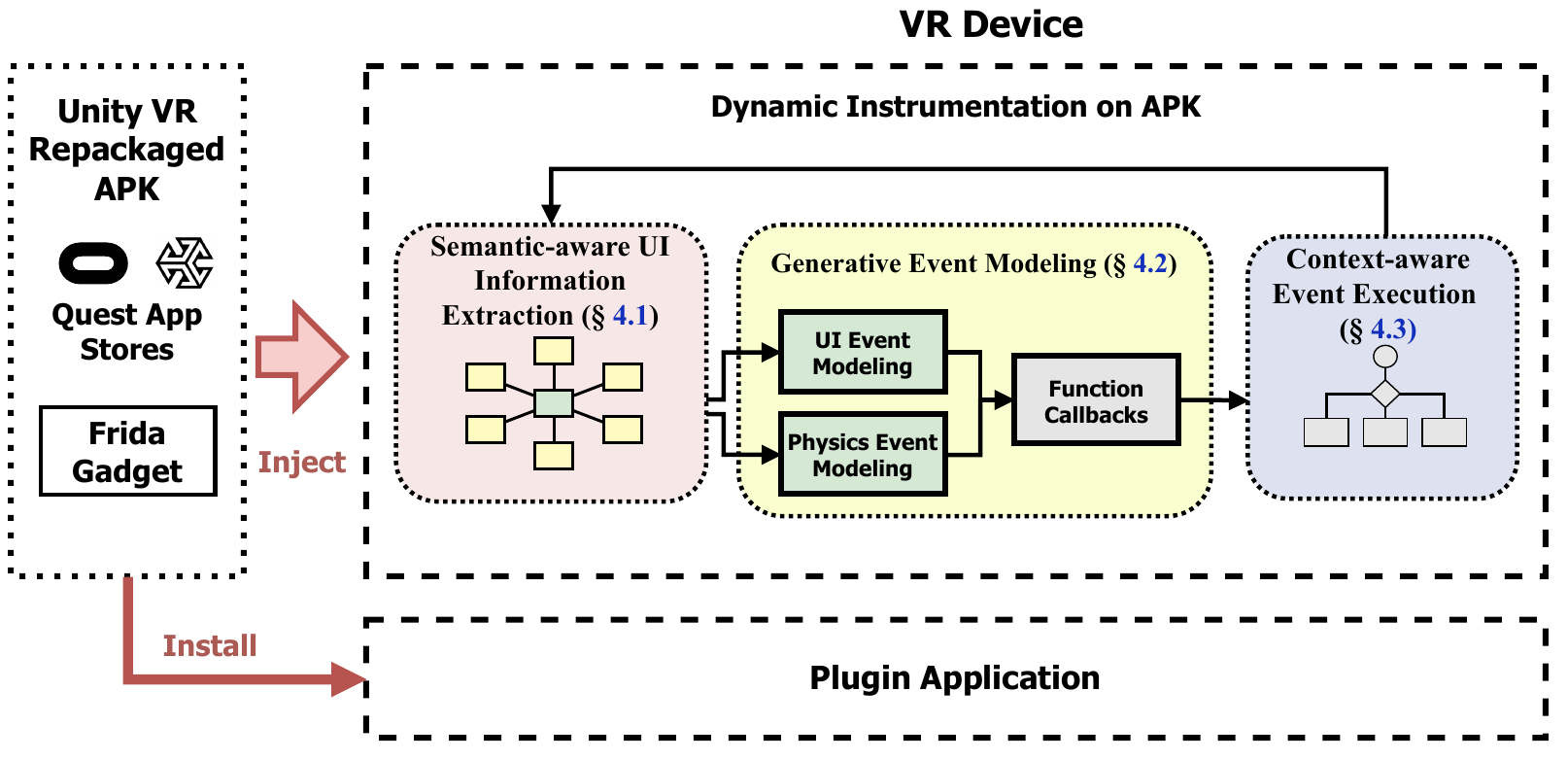}
          \vspace{-0.2in}
\caption{Overview of \sysname}
\label{fig:design}
\end{figure}

In this section, we present the detailed design of \sysname. We illustrate \sysname's overall design in \autoref{fig:design}. There are three key components inside \sysname: (1) UI Semantic Recovery (\S\ref{sec:semanticaware}), (2) Generative Event Modeling (\S\ref{sec:generative}), and (3) Context-aware Event Execution (\S\ref{sec:deps}). The final output is generated from the plugin application. In this work, our plugin application, \textit{AntMonitor}\cite{Le2015}, will produce information on sensitive data flows for each VR app tested. 

\subsection{UI Semantics Recovery}
\label{sec:semanticaware}
Identifying events requires knowledge of the UI semantics to eventually retrieve the EFCs within a scene. As such, to collect UI semantics, \sysname will first identify the UI components by extracting the class metadata (\S\ref{sec:classmeta}). Next, to identify the UI function callbacks, and subsequently all objects containing such callbacks, \sysname will then extract all available function metadata (\S\ref{sec:functionmeta}). Lastly, to identify which objects contain UI elements, \sysname will collect all objects within the loaded scene (\S\ref{sec:objectmeta}). With the class and function metadata, along with object information, \sysname eventually recovers the UI semantics of the loaded scene (\S\ref{sec:recovering}). \looseness=-1


\subsubsection{Extracting Class Metadata}
\label{sec:classmeta}
The \texttt{IL2CPP} runtime (i.e., \texttt{libil2cpp.so}) extracts and decrypts data from \texttt{global-metadata.dat}. More specifically, class name, function address, return type, and argument information can be extracted from the \texttt{IL2CPP} introspection APIs. In particular, we use APIs from the 9 functions prefixed with \texttt{il2cpp\_class} invoked by \textsf{Frida} to collect such metadata. These introspection APIs are detailed in \autoref{tab:il2cppapi} (see Appendix) for readers of interest. We notice that class metadata is stored as pointers in memory, and dereferencing such pointers can provide rich information from the class metadata, as well as the invocation of its own constructor, child methods, and fields. We also notice these class functions allow to invoke any child methods of the class, create a new instance, or access its field members. This functionality is especially important for UI information extraction, as class metadata is key to identifying UI elements within a VR app. \looseness=-1


\subsubsection{Extracting Function Metadata}
\label{sec:functionmeta}
Similarly to the class metadata, the function metadata is also stored in the \texttt{global-metadata.dat}. These function names are denoted using the combination of the function's residing class name along with the function name itself (e.g., in \autoref{fig:runex}: \texttt{Options\$\$OnOptionsClick}). We notice that the \texttt{IL2CPP} runtime initializes this metadata information once the app is started, storing it in memory. In order for the engine to properly invoke translated functions, and support reflection in C++, there must be generic C++ code to handle function signature extraction and invocation. This extraction is achieved by invoking C++ \texttt{IL2CPP} functions prefixed with \texttt{il2cpp\_method}. There are APIs for 7 such functions detailed in \autoref{tab:il2cppapi}. These functions return or handle references to every method offset from both C++ translation functions (i.e., \texttt{IL2CPP} functions) and translated C\# code (e.g., developer functions, and engine packages). \sysname essentially leverages these \texttt{IL2CPP} functions to extract function metadata, which will be used for EFC identification. By invoking such \texttt{IL2CPP} functions during the initial startup of the game, we create a global function table (GFT) with the entry address (i.e., the function offset) as the key, and the function's reference handle as the value. 

\subsubsection{Extracting Objects}
\label{sec:objectmeta}
To identify which objects are UI components and physics components, \sysname uses the \texttt{IL2CPP} runtime library to collect objects currently loaded in the scene using memory snapshots generated by \texttt{IL2CPP}. More specifically, we notice that the \texttt{IL2CPP} runtime can use memory snapshots to identify garbage collection handles that point to objects. This means we can invoke objects and the fields of such objects directly using \textsf{Frida}. Furthermore, we consider accessing field values and types essential for identifying UI events. Event objects are often stored within the fields of the residing object. For example, in \autoref{fig:runex}, we notice that \texttt{Options\$\$OnOptionsClick}, as well as nearly every UI \texttt{GameObject}, are attached with a Button component. We use two functions prefixed with \texttt{il2cpp\_object} in \autoref{tab:il2cppapi} to collect values of event objects. \looseness=-1

\subsubsection{Recovering \Unity UI elements}
\label{sec:recovering}
Recall in \S\ref{sec:background} that the UI elements are embedded in the \texttt{IL2CPP} app binary. As such, it is necessary for \sysname to access the \texttt{libil2cpp.so} binary and extract runtime UI objects which are in the form of \texttt{GameObjects} and Components. Using \textsf{Frida} with class metadata extraction as described, it is possible to extract the embedded \Unity UI element \texttt{GameObjects} by filtering for UI identifying classes. \Unity treats UI elements similarly to every other \texttt{GameObject} within the app. More specifically, the core developer scripts are built and managed using \texttt{GameObjects}, as such, \Unity treats UI elements as \texttt{GameObjects} that have attached UI components. For instance, in \autoref{fig:runex}, the \texttt{OPTIONS} button \texttt{GameObject} is a UI element because of the UI component attached to it. However, while it is possible to identify every UI component from the base \Unity SDK, developers may also implement their own UI components not derived from \Unity UI components.

\ignore{
\begin{figure}[t]
\centering
\scriptsize
 
\graphicspath{{./Figs/}}
\includegraphics[width=.35\textwidth, keepaspectratio]{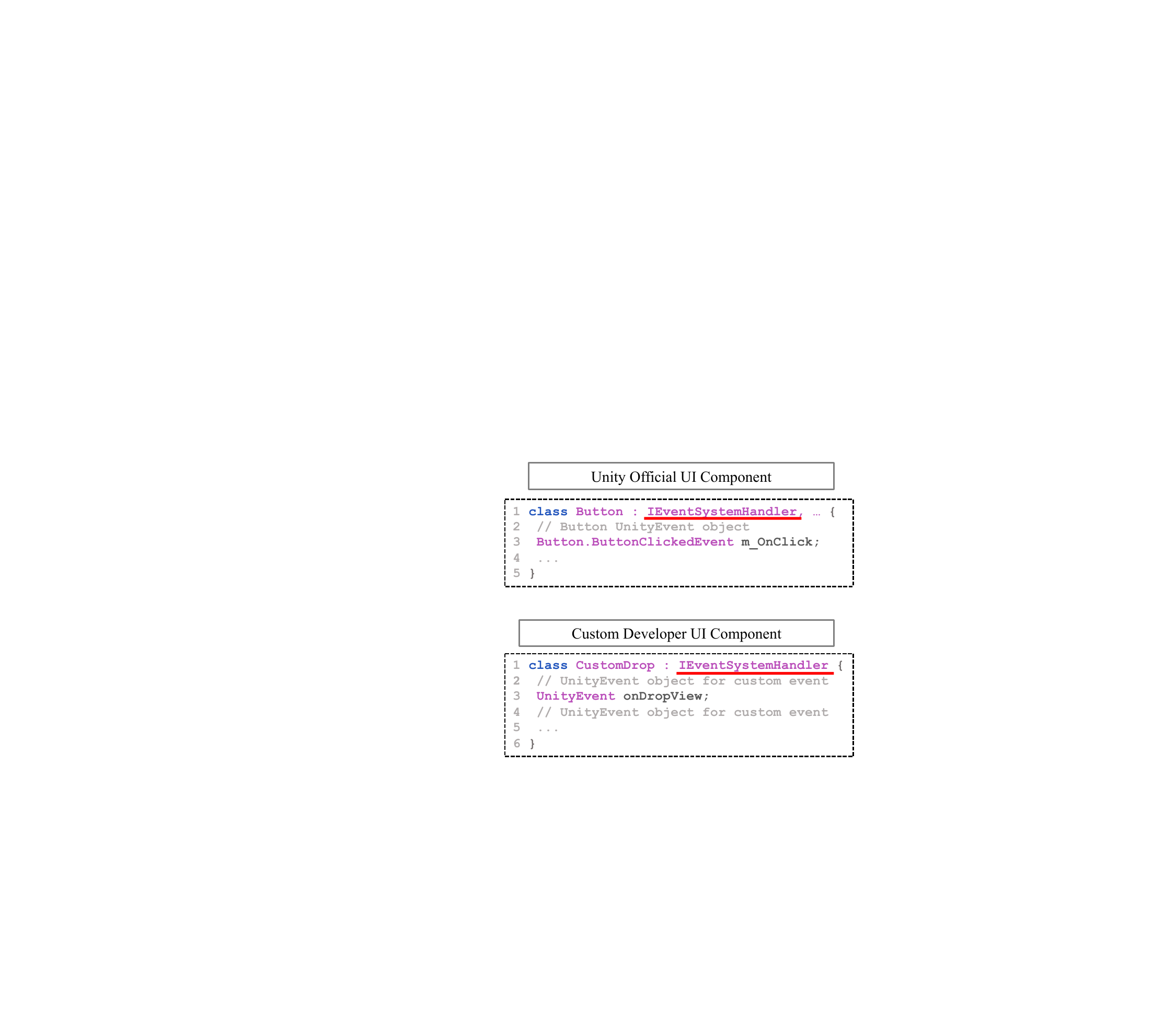}
\centering
\caption{UI identifying interface component view. \ZQ{is this figure refereed? how much value this figure will add?}}
\label{fig:uievents}
\end{figure}
}
We notice that the \Unity Engine describes all UI events under an event system interface. This interface must be implemented by every UI component, custom or not. We identify the core interface: \texttt{IEventSystemHandler} \cite{ieventsystemhandler}. In \autoref{fig:runex}, every UI button \texttt{GameObject} contains an attached Button component which its base class inherits the \texttt{IEventSystemHandler} interface. Using the class metadata extracted, \sysname identifies all used UI components by extracting the class inheritance hierarchy, and searching for the \texttt{IEventSystemHandler} interface.

\subsection{Generative Event Modeling}
\label{sec:generative}

To identify the event function callbacks (EFCs) in a VR app, \sysname must identify which objects such events reside. As described in \S\ref{sec:background} there are two types of \texttt{Unity} events: UI, and physics. Each event type must be extracted from the corresponding objects within an app scene. \Unity apps are dynamic. UI and physics events are resultant of the initialization of the scene and its \texttt{GameObjects}, unlike traditional event models which are hardcoded and static (e.g., Android UI XML model). \Unity UI however, will be generated as \texttt{GameObjects} are initialized either through scene initialization or event invocation. As such, \sysname will first identify all UI event objects from the loaded scene, and collect all EFCs from the UI event objects (\S\ref{sub:ui:events}). \sysname will then identify all available physics events and recover available triggerable and collisionable objects (\S\ref{sub:sub:physics:events}). \looseness=-1

\begin{figure}[t]
\centering
\scriptsize
 
\graphicspath{{./Figs/}}
\includegraphics[scale=.33, keepaspectratio]{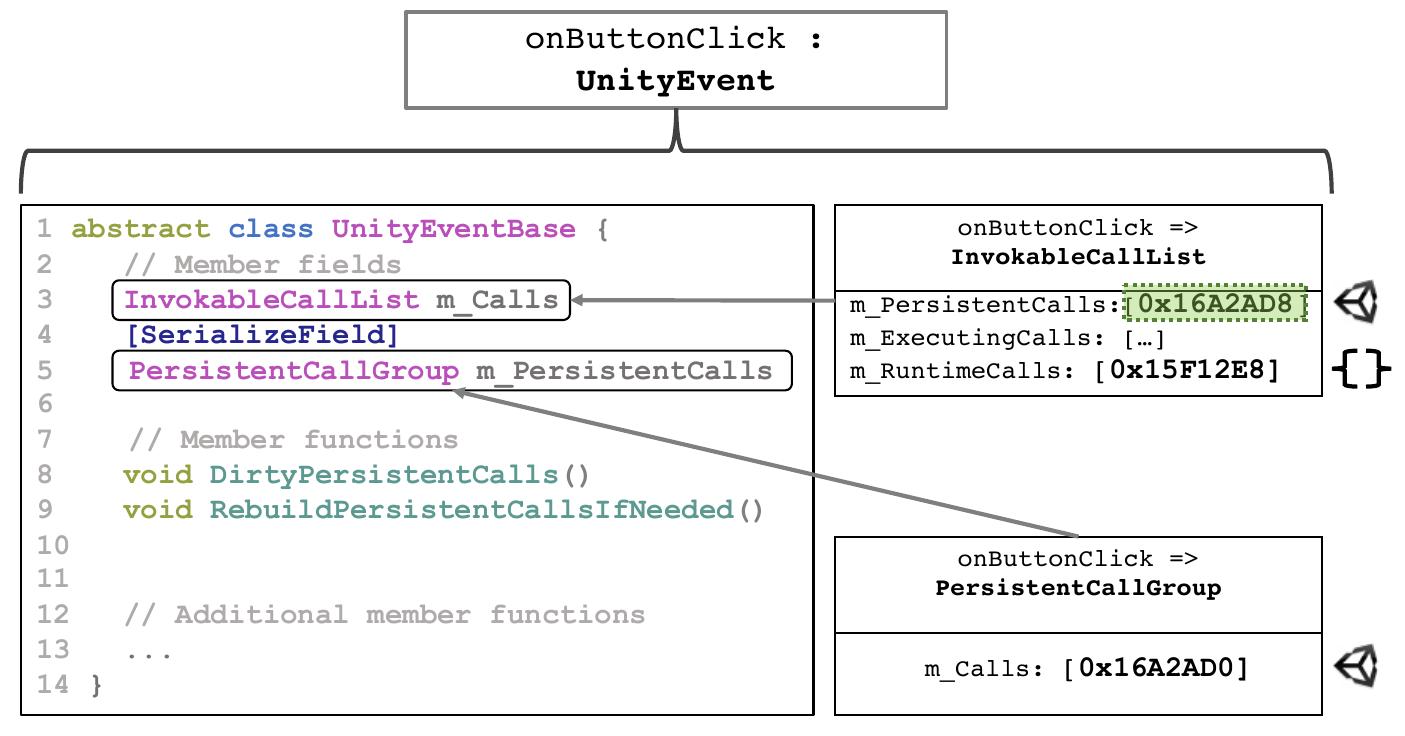}
\centering
\vspace{-0.1in}
\caption{Class structure of the UI-based \texttt{GameController} component from the running example. The (Unity) 3D box logo indicates that the function callback was attached via Unity Editor, while the bracket symbol indicates the function callback was attached via code. 
}
\label{fig:uievents}
\vspace{-0.3in}
\end{figure}

\subsubsection{Identifying UI Events}
\label{sub:ui:events}
After extracting and filtering the components described, \sysname then performs UI event identification. As mentioned in \S\ref{sec:background}, \texttt{Unity}'s UI SDK provides base components for developers to modify or inherit. These UI components store callbacks in the form of \texttt{UnityEvent} objects. \texttt{UnityEvent} objects are often stored within the fields of the residing component, or in the form of a derived class. From \autoref{fig:runex}, the \texttt{OPTIONS} Button component contains a \texttt{UnityEvent} called \texttt{onButtonClick} within the fields of the Button class.

Unfortunately, \texttt{UnityEvents} themselves are not the function callbacks that contain developer logic. \texttt{UnityEvents} are structured so that multiple function callbacks can be attached to one \texttt{UnityEvent} in a many-to-one representation. Furthermore, \texttt{UnityEvents} distinguish their function callbacks between runtime callbacks and persistent callbacks. Runtime callbacks are temporary callbacks that are added during the execution of the app through developer code. For example, in \autoref{fig:runex}, the \texttt{Start\$\$Press} EFC is assigned to the \texttt{START} Button via \Unity SDK API call: \texttt{UnityEvent.AddListener}. Persistent callbacks however, are serialized callbacks added prior to the execution of the app, typically within the \Unity Editor (e.g., \texttt{Options\$\$OnOptionsClick} in \autoref{fig:runex} and \autoref{fig:uievents}). We notice that persistent callbacks are not added through an API unless the player of the app invokes the corresponding event (e.g., through a button click). As such, API hooks alone will not be able to recover the persistent callbacks of the \texttt{UnityEvent}. \looseness=-1

Nevertheless, we notice that the base \texttt{UnityEvent} class, \texttt{UnityEventBase}, contains two fields: \texttt{m\_Calls}, and \texttt{m\_PersistentCalls}. These fields are of type \texttt{InvokableCallList} and \texttt{PersistentCallGroup}, respectively. Both of these fields store function callbacks into containers assigned to their respective \texttt{UnityEvent} as shown in \autoref{fig:uievents}. \texttt{InvokableCallList m\_Calls} stores the runtime callbacks and some persistent callbacks, while \texttt{PersistentCallGroup m\_PersistentCalls} stores the remaining persistent callbacks assigned by the developer at compile time. These callback containers contain the callback function metadata. This metadata provides the callback function's name, virtual address, argument cache, and the object target (i.e., the \texttt{UnityEvent} object). Using \texttt{\textsf{Frida}}, \sysname is able to extract the values of these callback containers, and thus, the callback functions themselves, using the functions from \autoref{tab:il2cppapi}. \looseness=-1

\subsubsection{Identifying Physics Events}
\label{sub:sub:physics:events}
As described in \S\ref{sec:background}, there are two types of physics events: \textbf{collisions}, and \textbf{triggers}. Fortunately, physics events are slightly easier to identify than UI events, as the EFCs are not hidden within field objects. However, physics events must follow specific rules dictated by the game engine, these rules are shown in \autoref{tab:rulematrix}. There are three important Collider properties to help identify physics interactions, Rigidbody attached, Static property, and Kinematic property~\cite{rulematrix}. Rigidbodies can be attached to a Collider component to apply physics motions, static property indicates the collider is a non moving object but allows physics events to take place, while the kinematic property indicates the collider behaves static but also allow the movement of the collider object. In the following, we describe how \sysname identifies and extracts such collider objects to execute physics events.

\begin{table}[t]
\begin{minipage}{.45\linewidth}
\scriptsize
\setlength\extrarowheight{2pt} 
\begin{tabular}{|c|c|c|c|c|c|c|}
  \hline
  & \multicolumn{3}{|c|}{Trigger Colliders} &  \multicolumn{3}{|c|}{Collision Colliders}\\ \hline
     & RB & Static & Kinematic & RB & Static & Kinematic \\ 
  \hline
  RB & \ding{55} & \ding{55} & \ding{55}  & \ding{51} & \ding{51} & \ding{51} \\
  \hline
  Static & \ding{55} & \ding{55} & \ding{51}& \ding{51} & \ding{55} & \ding{55}  \\
  \hline
  Kinematic & \ding{55} & \ding{51} & \ding{51} & \ding{51} & \ding{55} & \ding{55}  \\
  \hline
\end{tabular}
\end{minipage}

\ignore{
\begin{minipage}{.45\linewidth}
\scriptsize
\setlength\extrarowheight{2pt} 
\begin{tabular}{|c|c|c|c|}
  \hline
  \multicolumn{4}{|c|}{Collision Colliders} \\
  \hline
   & RB & Static & Kinematic \\
  \hline
  RB & \ding{51} & \ding{51} & \ding{51}  \\
  \hline
  Static & \ding{51} & \ding{55} & \ding{55}  \\
  \hline
  Kinematic & \ding{51} & \ding{55} & \ding{55}  \\
  \hline
\end{tabular}
\end{minipage}
}
\caption{Rule matrix of two colliders executing physics events where \ding{51} indicates event is executable, and \ding{55} not executable. RB = colliders with Rigidbody attached, Static = colliders without a Rigidbody, Kinematic = RB colliders with the kinematic property set to true. 
}
\label{tab:rulematrix}
\vspace{-0.1in}
\end{table}

\paragraph{Trigger Events} A trigger event occurs when a \textbf{non-solid} game object (i.e., a game object that does not contain a \texttt{Rigidbody} component) intersects with other non-solid game objects. We call these non-solid game objects \textbf{triggerables}. There are three properties that determine if a game object is an invokable triggerable:
\begin{packeditemize}
     \item [\textbf{(1)}]  A collider is attached to the game object.
    \item [\textbf{(2)}] The collider component must have \texttt{isTrigger} set to \textbf{true}.
    \item [\textbf{(3)}] At least one of the game object's components implements at least one of the following trigger functions: \texttt{OnTriggerEnter}, \texttt{OnTriggerStay}, and \texttt{OnTriggerExit}.
\end{packeditemize}
To identify other collider objects to invoke these triggers with, we present the Trigger Collider matrix from the 2nd to the 4th column in \autoref{tab:rulematrix}. Collider components may contain a Rigidbody to allow physics movements to be applied to its host GameObject. This is also how we identify `solid' objects as mentioned in \S\ref{sec:background}. However, trigger events are simply intersections between Collider components that do not contain any physics movement properties. As such, only colliders without a Rigidbody attached may be used with trigger events.

To identify all invokable triggerable game objects, \sysname uses the components described in \S\ref{sec:objectmeta} to filter out components that do not have the three properties, and extract the virtual addresses of the trigger functions. These virtual addresses are then sent to our {\it Dependence Resolution} (\S\ref{subsub:dr}) to resolve dependencies. Furthermore, to identify which Collider components may interact with such triggerable events, we filter out components using the Trigger Collider rule matrix from \autoref{tab:rulematrix}. \looseness=-1

\paragraph{Collision Events} A collision event occurs when a \textbf{solid} game object (i.e., game object contains a \texttt{Rigidbody} component) collides with other solid game objects. We call these solid game objects \textbf{collisionables}. There are four properties that determine if a game object is an invokable collisionable:
\begin{packeditemize}
\item [\textbf{(1)}]  A collider is attached to the game object.
    \item [\textbf{(2)}] A \texttt{Rigidbody} component is attached to the game object.
    \item [\textbf{(3)}] The collider component must have \texttt{isTrigger} set to \textbf{false}.
    \item [\textbf{(4)}] At least one of the game object's components implement at least one of the following collision functions: \texttt{OnCollisionEnter}, \texttt{OnCollisionStay}, and \texttt{OnCollisionExit}.
\end{packeditemize}
To identify other collider objects to invoke these collision events with, we present the Collision Collider matrix from the 5th to the 7th column in \autoref{tab:rulematrix}. However, unlike trigger colliders, at least one of the two intersecting colliders must have a Rigidbody attached to it. As a result, \sysname will filter for collider objects that follow the Collision Collider matrix when invoking the current invokable collisionable function, and  extract the virtual addresses of these collision functions for the follow-up analysis. \looseness=-1


\subsection{Context-aware Event Execution}
\label{sec:deps}

Once \sysname collects the initial events, the execution can be done using the extracted function offset and invocation by \textsf{Frida}. However, as outlined in \S\ref{sec:chall}, the execution of events may cause more events to be enabled and/or other events to be disabled. These dependencies depicted in \autoref{fig:deps} demonstrate that when the \texttt{OPTIONS} button is clicked, the \texttt{START} and \texttt{QUIT} buttons become disabled, while the \texttt{TurnOptions}, \texttt{VolumeOptions}, and \texttt{ValidateOptions} buttons are enabled. To resolve such dependencies from event execution, \sysname will first execute the first found initial event from the loaded scene, then once executed, \sysname will attempt to find dependencies from the new state (\S\ref{subsub:eedi}). Subsequently, \sysname will then resolve the identified dependencies if there are any (\S\ref{subsub:dr}).


\subsubsection{Event Execution and Dependency Identification}
\label{subsub:eedi}
To cover and execute EFCs, \sysname performs context-aware event execution by analyzing dynamic dependencies of all identified EFCs collected. More specifically, once all the EFCs are extracted, execution of events can be performed through direct invocation using \textsf{Frida} along with the GFT to extract the function offset and parameters. To find object parameters, we retrieve all currently available objects (collected upon scene initialization) to find any active objects to test the EFC with. At this stage, possible plugin applications such as network traffic collection applications (e.g., AntMonitor \cite{Le2015}) can be used to observe the effects of executing the identified EFCs. Initially, the first EFC is chosen at random; however, as mentioned in \S\ref{sec:chall} and illustrated in \autoref{fig:deps}, it is possible that executing the EFC causes other EFCs to be enabled or disabled. In such a case, we identify these newly changed or added EFCs as dependencies, and map a tree of events to model their dependencies. \looseness=-1

\subsubsection{Dependency Resolution}
\label{subsub:dr}
To cover the entire tree from \autoref{fig:deps}, \sysname will maintain a state machine, identifying every new event linked with its parent. An example is the recovery of the initial state to recover the disabled events in \autoref{fig:deps}. There are two cases of recovering state:
\begin{packeditemize}
    \item The parent EFC is a root event of the current scene. In this case, recovering the initial state is necessary to recover disabled events. As such, \sysname will reload the current scene to recover the disabled callbacks and execute on a different event path.
    \item The parent EFC is a child event of the current scene. In this case, to recover the events generated by this parent EFC, \sysname invokes the parent EFC again without reloading the scene. Therefore, the disabled events in this level will be recovered and callbacks will execute on a different path.
\end{packeditemize}

Using these recovery methodologies, the event dependencies will be resolved. This way, the sequence of events can be covered holistically, and all dependencies can be resolved. As many VR apps, such as video games where the state of the game is important to covering every execution path, we must be sure to cover dependencies and the sequence of events to resolve such dependencies.

\section{Implementation}
\label{sec:implementation}

\vspace{-0.1in}
\noindent \textbf{Dynamic Instrumentation.} We have implemented a prototype of \sysname atop \textsf{Frida} as our main dynamic instrumentation engine. \textsf{Frida} contains multiple modules with both JavaScript and Python bindings. In particular, we use \textsf{Frida}'s \texttt{Interceptor}~\cite{interceptor} module to perform API hooks on function addresses, \texttt{NativeFunction} to perform invocation of such function addresses, and \texttt{Instruction} module to reassemble ARM64 instructions from such function addresses. We have developed \sysname with over 3,000 lines of TypeScript code and over 1,000 lines of Python code. \looseness=-1

To perform dynamic instrumentation to collect and invoke events, \sysname uses \textsf{Frida} as the dynamic instrumentation toolkit and interfaces with the functions shown in \autoref{tab:il2cppapi}.
\textsf{Frida} is the only dynamic instrumentation toolkit and supports non-rooted Android devices, which is crucial as Quest 2 devices are typically root locked. 
Therefore, for non-rooted Quest devices, we must inject the \textsf{Frida} server binary into every tested app. We use \texttt{frida-gadget}~\cite{gadget}, a \textsf{Frida} server binary that can be injected directly into the game source files. We then use \texttt{objection}~\cite{sensepost} to repackage the APK with the injected \textsf{Frida} server binary, and install the modified APK into the Quest 2 device using \texttt{adb}.


\paragraph{Application: Sensitive Data Detection}
\sysname supports various security applications, including crash detection and malware analysis, but this work focuses on detecting sensitive data exposure. As an event exploration and dynamic analysis framework, \sysname is application-agnostic. For sensitive data detection, \sysname employs an adapted version of \textit{AntMonitor}~\cite{Le2015} to intercept outgoing TLS traffic from VR apps. However, SSL pinning in third-party \Unity apps presents challenges, as it encrypts network traffic and obscures potential exposures.

To address this, \sysname integrates \textit{AntMonitor} for TLS traffic capture and employs decryption techniques to bypass SSL pinning in both Android and Unity layers. While bypassing SSL pinning in Android apps is well-documented~\cite{Antonishyn2020}, Unity apps introduce additional challenges. Older Unity versions use the \texttt{mbedtls\_x509\_crt\_verify\_with\_profile} function for verification~\cite{277092}, but apps released after 2021 often utilize \texttt{x509\_crt\_verify\_restartable\_ca\_cb}. To handle this, \sysname supports bypassing the new verification function using Frida for API hooking, nullifying flags and return values.

During experiments, \textit{AntMonitor} collected network traffic while \sysname triggered sensitive data flows through automated event execution, unlike prior works like OVRseen~\cite{277092}, which relied on human-triggered events. While \sysname enables the discovery of otherwise hidden data flows, not all such flows constitute privacy leaks, as developers may disclose them in privacy policies.

\section{Evaluation}
\label{sec:evaluation}

In this section, we present the evaluation results of \sysname. First, we outline our experiment setup (\S\ref{sub:setup}), followed by its effectiveness including the capability for detecting privacy data exposure to current dynamic analysis tool Android Monkey (\S\ref{sub:effectiveness}). Finally, we present the efficiency overhead of \sysname (\S\ref{sub:performance}). 

\subsection{Experiment Setup}
\label{sub:setup}

\begin{table*}[t]
\scriptsize
\centering
\begin{tabular}{|l||c|c|c|c|c|c|c|c|c|c|c|c|c|}
\hline
\multicolumn{13}{|c|}{\it{Scene 0}} \\
\hline
{\bf{Event ID}} &  {\bf 1} & {\bf 1a} & {\bf 2} & {\bf 2a} & {\bf 2b} & {\bf 2c} & {\bf 2d} & {\bf 3} & {\bf 3a} & {\bf 3b} & {\bf 3c} & {\bf 4} \\ \hline

\it{\sysname}&41&40&41&40&39&40&39&41&40&39&39&41 \\ \hline
\it{Monkey}&0&1&0&0&0&0&0&0&0&0&0&0 \\ \hline
\multicolumn{13}{|c|}{\it{Scene 1}} \\
\hline

{\bf{Event ID}} & {\bf CCube\#1} & {\bf  CCube\#2} & {\bf  CCube\#3} & {\bf  CCube\#4} & {\bf  CCube\#5} & {\bf  CCube\#6} & {\bf  CCube\#7} & {\bf  CCube\#8} & {\bf  CCube\#9} & {\bf  TCube\#1} & - & - \\ \hline

\it{\sysname} & 47 & 212 & 355 & 249 & 243 & 200 & 108 & 619 & 503 & 9 & - & -\\ \hline
\it{Monkey} & 0 & 0 & 0 & 0 & 0 & 0 & 0 & 0 & 0 & 0 & - & -\\ \hline

\multicolumn{13}{|c|}{\it{Scene 2}} \\
\hline

{\bf{Event ID}} & {\bf 1} & {\bf  2} & {\bf  3} & {\bf  4} & {\bf  5} & {\bf  6} & {\bf  7} & {\bf  CCube\#1} & {\bf  CCube\#2} & {\bf  TCube\#1} & {\bf  TCube\#2} & -   \\ \hline

\it{\sysname}&19&19&19&4&18&3&17&133&133&10&10 & -  \\ \hline
\it{Monkey}&2&0&1&0&0&0&0&0&0&0&0 & - \\ \hline

\end{tabular}
\vspace{-0.1in}
\caption{Total number of events triggered by \sysname and \texttt{Monkey}, grouped by scene number from \autoref{fig:model}. Physics events (e.g., CCube\#1, TCube\#1) sum up the total events triggered from all three callbacks (e.g., \texttt{On(Trigger/Collision)Enter}, \texttt{On(Trigger/Collision)Stay}, \texttt{On(Trigger/Collision)Exit}). \color{black}}
\label{tab:modelresults}
\end{table*}

\vspace{-0.1in}

\begin{figure}[t]
    \centering
        \includegraphics[width=\linewidth]{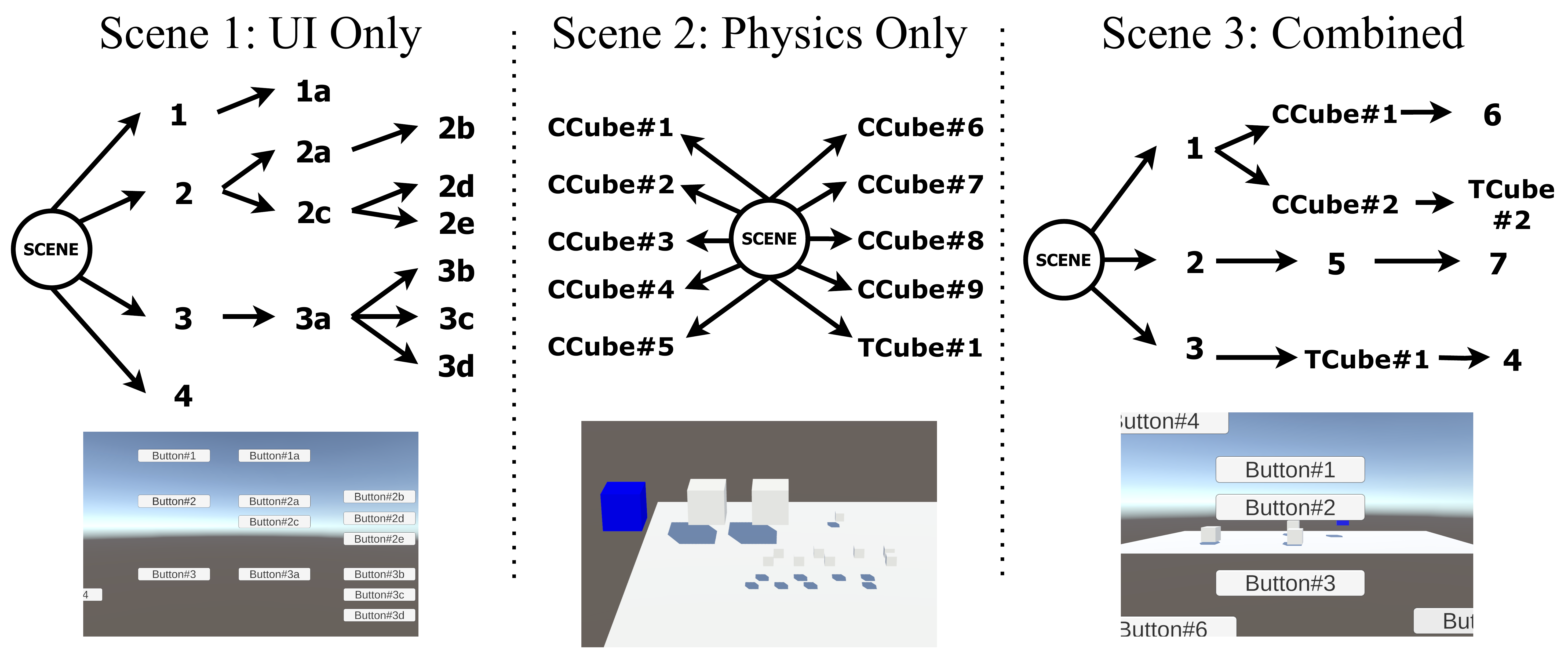}
\caption{Custom VR app describing the dependency structure for each scene, where each alphanumerical value (e.g., 1, 2d, 3a, etc.) indicates a UI event, each "Cube" prefixed with "C" indicates a collisionable event and "T" indicates a triggerable event. }
\vspace{-0.2in}
\label{fig:model}
\end{figure}

\paragraph{A Custom VR Unity App} 
To show the effectiveness of \sysname's event exploration and execution capabilities, we have developed a custom Unity VR app that contains both UI and physics events and their dependencies. To accurately test event exercising, we develop the app by ourselves so that we know exactly the number of scenes, events, and UIs and we can also easily instrument the app to log the event exercise behaviors.  
In particular, within the app, we developed three scenes: (1) only UI events, (2) only physics events, (3) a combination of UI and physics events. To show the dependency structure of such scenes, we present \autoref{fig:model}, where a screenshot of each scene is shown with the events enabled to clearly illustrate the structure. Additionally, to test whether \sysname can execute an event beyond the screen, we placed event 4 from scene 1, and events 4 and 6 from scene 3, outside the default field of view.
\color{black}

\bheading{Unity game acquisition}. To test \sysname across the VR app ecosystem, w\color{black}e collected 263 free games in total from both the Meta Quest app store and the SideQuest app store. SideQuest is a third-party endorsed app store, typically where developers publish apps or games for facilitating distribution or for early access releases. Specifically, we have scraped 84 games from the free Meta Quest store and 179 games from the SideQuest store. Additionally, we collected 103 paid games from the Meta Quest app store, totaling 366 games. Scraping Meta Quest and SideQuest resulted in metadata acquisition from each individual game collected. Such metadata includes app rating, number of downloads, category keywords, and privacy policy links. \looseness=-1 
\color{black}

\bheading{Experiment Environment}.
\sysname was deployed in an Ubuntu 21.10 LTS desktop environment, running on a machine equipped with an Intel i7-8750H processor, 4 CPU cores, and 16 GB of RAM.
While more processing power could be allocated for \sysname, the primary bottleneck in our tests was the event modeling and execution process, which runs on the Quest 2 devices. The 4 used Quest 2 devices runs on 8 CPU cores from a Qualcomm Snapdragon XR2 processor and includes 6 GB of RAM~\cite{moody_2021}.\looseness=-1

\subsection{Effectiveness}
\label{sub:effectiveness}

\begin{table*}[t]
\scriptsize
\centering
\begin{tabular}{l|r|r|r|r|r|r|r||r|r|r|r|r|r|r}
\hline
 & \multicolumn{7}{c||}{Free Apps}  & \multicolumn{7}{c}{Paid Apps} \\ \cline{2-15}

{\bf Rating} &  \rot{\bf \# Apps} & \rot{\bf\# Scenes} & \rot{\bf\# GameObject} & \rot{\bf\# UI Events} & \rot{\bf\# Collisions} & \rot{\bf\# Triggers} & \rot{\bf Time (s)} & \rot{\bf \# Apps} & \rot{\bf\# Scenes} & \rot{\bf\# GameObject} & \rot{\bf\# UI Events} & \rot{\bf\# Collisions} & \rot{\bf\# Triggers} & \rot{\bf Time (s)}\\ \hline

[2.5, 2.75)&23&57&61,357&913&494&775&3353.20&3&4&1,536&0&0&13&205.21 \\ \hline
[2.75, 3.0)&2&14&9,778&28&0&147&175.89&1&28&7,722&0&0&54&472.36 \\ \hline
[3.0, 3.25)&6&18&35,914&82&780&281&1956.54&1&11&6,826&5&0&2&243.72 \\ \hline
[3.25, 3.5)&9&17&29,162&2,896&0&7&1176.12&3&48&42,789&0&0&2&807.12 \\ \hline
[3.5, 3.75)&20&99&65,731&11,320&170&541&3696.56&5&17&12,016&99&2&877&861.57 \\ \hline
[4.0, 4.25)&39&158&158,823&10,879&191&3116&17360.33 &14&45&66,106&130&230&266&3126.51 \\ \hline
[4.25, 4.5)&39&158&222,434&6,474&1139&2774&10238.70&18&30&49,873&1,483&0&107&3111.51 \\ \hline
[4.5, 4.75)&41&187&327,299&1,999&1114&4082&5811.47&29&194&210,346&4,396&451&520&6590.65 \\ \hline
[4.75, 5)&84&426&439,990&22,118&4967&5466&16773.521&29&217&115,270&1,338&341&632&5437.94 \\ \hline

\end{tabular}
\vspace{-0.1in}
\caption{Aggregated data for Meta apps \& SideQuest apps based on ratings, separating paid and free games.}
\vspace{-0.1in}
\label{tab:meta:side:quest}
\end{table*}

\vspace{-0.1in}
\paragraph{Custom VR App Verification}
We first verified that \sysname effectively executes and accurately identifies scenes, \texttt{GameObject}s, as well as UI, collision, and trigger events within the custom VR app. The goal state being, \sysname, traverses all events of each scene.

We present the results of \sysname running on the custom VR app in \autoref{tab:modelresults}. We first notice a sizable number of collision events triggered. Specifically, for each collision event, the \texttt{OnCollisionStay} EFC is vastly higher than the \texttt{OnCollisionEnter/Exit} EFC. Due to the necessity of moving collider components to different positions to trigger collision events, a short delay (300 ms) is introduced between collisions to ensure collision event invocation. \texttt{OnCollisionStay} follows the \texttt{FixedUpdate} time rate (i.e., 20 ms per call~\cite{unityexecutionorder, fixedupdate}), as such, each collision event will call \texttt{OnCollisionStay} at least 15 times (300ms / 20ms). Additionally, for each collision event, \sysname will attempt to collide all collisionable \texttt{GameObjects} with the acting collider within the scene, further bloating the occurrences.

Futhermore, as shown in \autoref{tab:modelresults}, \sysname was able to comprehensively invoke all given UI and physics events while also traversing the internal scenes. \sysname was able to invoke events within a nested state (i.e., triggered all events with dependencies), as well as explore the dependencies of interleaving physics and UI events. Additionally, \sysname was able to invoke events 4 (scene 1), 5 and 6 (scene 2), which are \textbf{outside} the field-of-view of the default headset position. Note that all invocations are recorded by the custom app. Because \sysname attempts to trigger an event on all available threads, it is highly likely that \sysname invokes the same event multiple times per execution. Moreover, we notice fewer invocations the deeper the UI event lies within the dependency tree (e.g., events 4 and 6 from scene 2). The same initial events will be invoked more times than the nested ones due to the backtracking behavior described in \S\ref{subsub:dr}. \looseness=-1

\paragraph{VR Apps from App Stores} Next, we ran \sysname on the corpus of 366 VR apps collected. \color{black} The aggregated results for this experiment are presented in~\autoref{tab:meta:side:quest}. Specifically, we aggregated the runtime data results based on the app's rating listed on the MetaQuest and SideQuest stores. The metadata from the SideQuest and Meta Quest stores contain largely inconsistent data keys, with some intersection, specifically the user ratings, as such, we use ratings as the primary key to describe the experiment results. \color{black} We observe that the category comprising apps with ratings ranging from 4.75 to 5 (denoted as [4.75, 5)) represents the largest subset within our dataset and, consequently, contains the highest number of \texttt{GameObject}s. The table also reveals a direct correlation between the number of \texttt{GameObject}s and the number of both UI and physics events (collisions and triggers). This correlation is expected, as events are inherently tied to \texttt{GameObject}s within a scene. Therefore, apps with a greater number of \texttt{GameObject}s are likely to feature more event-associated \texttt{GameObject}s. 
While a noticeable correlation exists between the number of scenes and the number of \texttt{GameObject}s, the latter is primarily influenced by the developer's design choices within each scene. For example, the paid apps under [4.5, 4.75) range contains more \texttt{GameObject}s than paid apps under [4.75, 5), however, the scene count for [4.75, 5) is larger than the [4.5, 4.75) range. The same can be said for [4.0, 4.25) and [4.25, 4.5) paid apps.

\begin{figure*}[t]
    \centering
        \centering
        \vspace{-0.1in}
        \includegraphics[width =.8\linewidth]{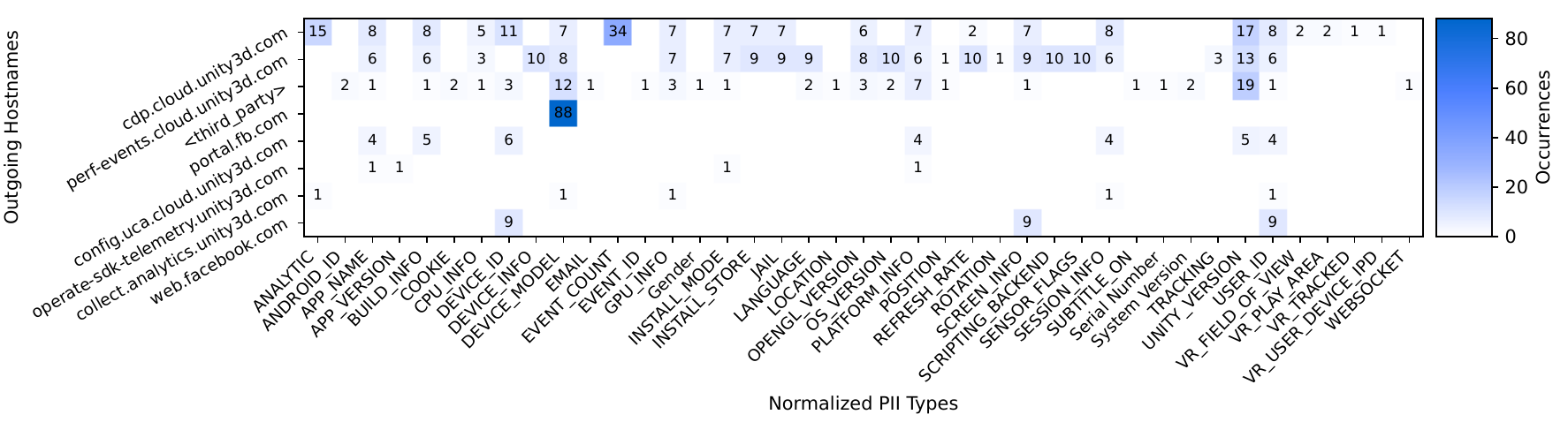}
        \label{fig:hostnameheat}
     \vspace{-0.1in}
\caption{A heatmap showing unique occurrences of sensitive data flow types per app rating range of both free and paid Meta Quest apps and SideQuest apps. }
\label{fig:ratingleaks}
\end{figure*}

\begin{figure*}[t]
    \centering
        \centering
        \vspace{-0.1in}
        \includegraphics[width=.8\linewidth]{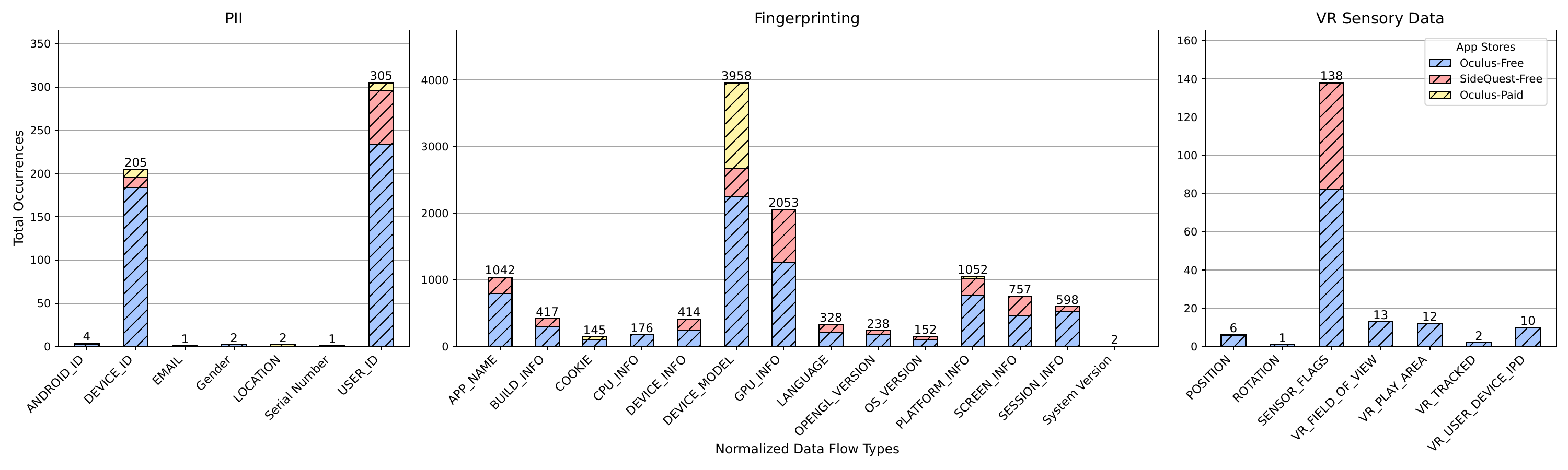}
        \label{fig:total_flows}
     \vspace{-0.2in}
\caption{Total number of sensitive data flow occurrences grouped by app store, categorized by PII (Personal Identifiable Information), Fingerprinting, VR Sensory Data.}
\label{fig:total_flows}
\end{figure*}

\begin{figure*}[t]
    \centering
        \centering
        \vspace{-0.1in}
        \includegraphics[width=.8\linewidth]{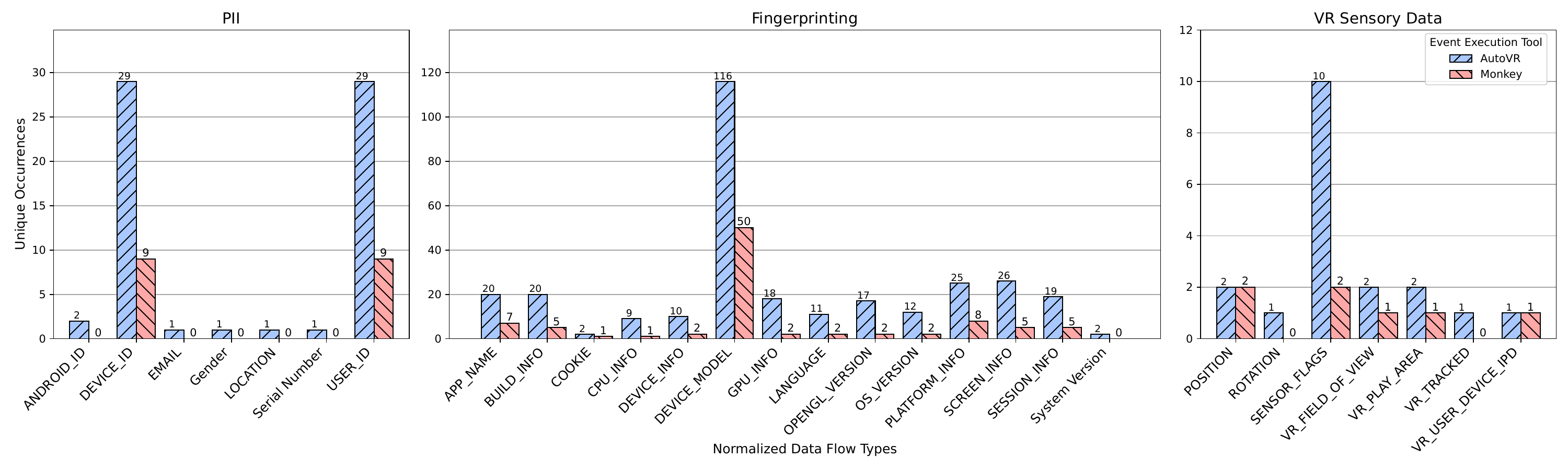}
        \label{fig:comparemonkey}
     \vspace{-0.2in}
\caption{Comparison of the effectiveness with respect to the number of unique sensitive data flow occurrences found using \sysname vs Monkey.}
\label{fig:comparemonkey}
\vspace{-0.15in}
\end{figure*}


From both tables, we also notice a direct correlation between the number of collision and trigger events to the amount of time taken for execution. Due to the necessity of moving collider components to different positions to trigger collision events, a short delay (300 ms) is introduced between collisions to ensure collision event invocation. Otherwise, the game engine will not be able to handle too many collisions in a short period of time, causing the game to crash. As collider components are highly prevalent in VR apps, which will produce more collisions and triggers, the time taken to execute \sysname on VR apps is significant.
\looseness=-1
\paragraph{Sensitive Data Exposure Detection} Next, to quantitatively evaluate the effectiveness of applying \sysname for sensitive data exposure detection, we showcase \autoref{fig:total_flows}. The X-axis represents the types of sensitive data exposures, while the Y-axis indicates the number of occurrences found in our total corpus of VR apps. Unsurprisingly, the \textsf{Unity} version is the most popular data exposed by all the apps, which is typically used for game metrics sent to the developer. The next three top data types (i.e., \texttt{APP\_INFO}, \texttt{PLATFORM\_INFO}, \texttt{SESSION\_DATA}) are potentially used for digital fingerprinting~\cite{277092}. Additionally, data types such as \texttt{SCREEN\_INFO}, \texttt{GPU\_INFO}, \texttt{CPU\_INFO}, and \texttt{DEVICE\_INFO} are collected to increase fingerprinting entropy, commonly used to accurately track user behavior and identifiablility~\cite{bacis2024assessing, 277092}. Throughout our experiments however, we noticed that \texttt{USER\_ID} and \texttt{DEVICE\_ID} are the most stable identifier amongst the VR apps, and is one of the common data points we see in \autoref{fig:ratingleaks}. 

We notice from \autoref{fig:total_flows} that there are large discrepancies in data flows for free and paid apps (e.g., USER\_ID and DEVICE\_ID). Many of the paid apps were generally not exposing sensitive data, as we collected exposures from only 56 apps. Additionally, we noticed application-side encryption in many of the paid app's network packets. As these could not be automatically decrypted beyond just SSL/TLS decryption, much of the data could not be interpreted. We understand that paid apps are generally well built, as most of the paid apps are within the 3.5 rating+ range (see \autoref{tab:meta:side:quest}), as such, application-side encryption would likely be integrated into the apps. Free apps, conversely, generally rely on the free in-house tools provided by Unity to send network traffic, and many apps have integrated Unity’s in-house analytics to send data regarding user behavior (e.g., in \autoref{case:study2}, we notice such analytical traffic where ``userid" and ``deviceid" data pairs are being sent). The two top outgoing hostnames (\textit{cdp.cloud.unity3d.com} and \textit{perf-events.cloud.unity3d.com}), are specifically used for analytics and performance. Additionally, there are more free apps in our corpus than paid apps, further inflating the number of data exposures found.
\color{black}

Furthermore, to quantify the correlation of the total unique sensitive data flows to the destination hostname, we present \autoref{fig:ratingleaks}. We notice that majority of the data flows reach a \textit{unity3d.com} domain, Facebook domain (i.e., \textit{fb.com} and \textit{facebook.com}). As such, we grouped the third-party domains into one entity. We denote 3rd party domains as any domain not associated with Meta, Facebook, or Unity. We identify that significant traffic of analytic and device data outgoing to \textit{cdp.cloud.unity3d.com} and \textit{perf-events.cloud.unity3d.com}. \textit{perf-events.cloud.unity3d.com} is typically used for social features, where analytical data is needed to support such features\cite{277092}. Additionally, other network traffic outgoing to \texttt{*.cloud.unity3d.com} could be used by the developer to obtain analytical or device information. We show an example network packet in \autoref{case:study2} in the Appendix.  \looseness=-1

\paragraph{Comparison with Monkey}
To compare the results of \sysname with \textsf{Monkey}, we performed the same data collection process used for \sysname to \textsf{Monkey}, using \textit{AntMonitor} to collect network data and \textsf{Frida} for event trigger data collection. We allotted \textsf{Monkey} 20 minutes of running time and set \sysname to a 20 minute timeout to fairly assess the comparison.  \looseness=-1

We present a comparative analysis of the detected sensitive data exposure by \textsf{Monkey} and \sysname in \autoref{fig:comparemonkey}. Both tools were ran against the 366 apps. \sysname has collected 390 \textbf{unique} data flow exposures, whereas \textsf{Monkey} has gathered only 117. This represents a 2.2x improvement in finding \textbf{unique} sensitive data flow exposures.

Uniquely, \sysname is capable of capturing more unique PII data flows, such as \texttt{EMAIL, Gender, LOCATION, and Serial Number}. Uninspiringly, \sysname was able to trigger more sensitive data flow exposures, as \sysname's context-aware execution allow the game to enter more states as more events are triggered. Furthermore, because \sysname also traverses through the app scenes, which allows \sysname to load new events in different levels that \texttt{Monkey} cannot access. Therefore, \sysname will visit more \texttt{GameObjects}, and trigger both 3D collision/trigger and UI events, leading to more outgoing data flows. 
\looseness=-1

\ignore{
\begin{figure}
    \centering
        \includegraphics[width=.75\columnwidth]{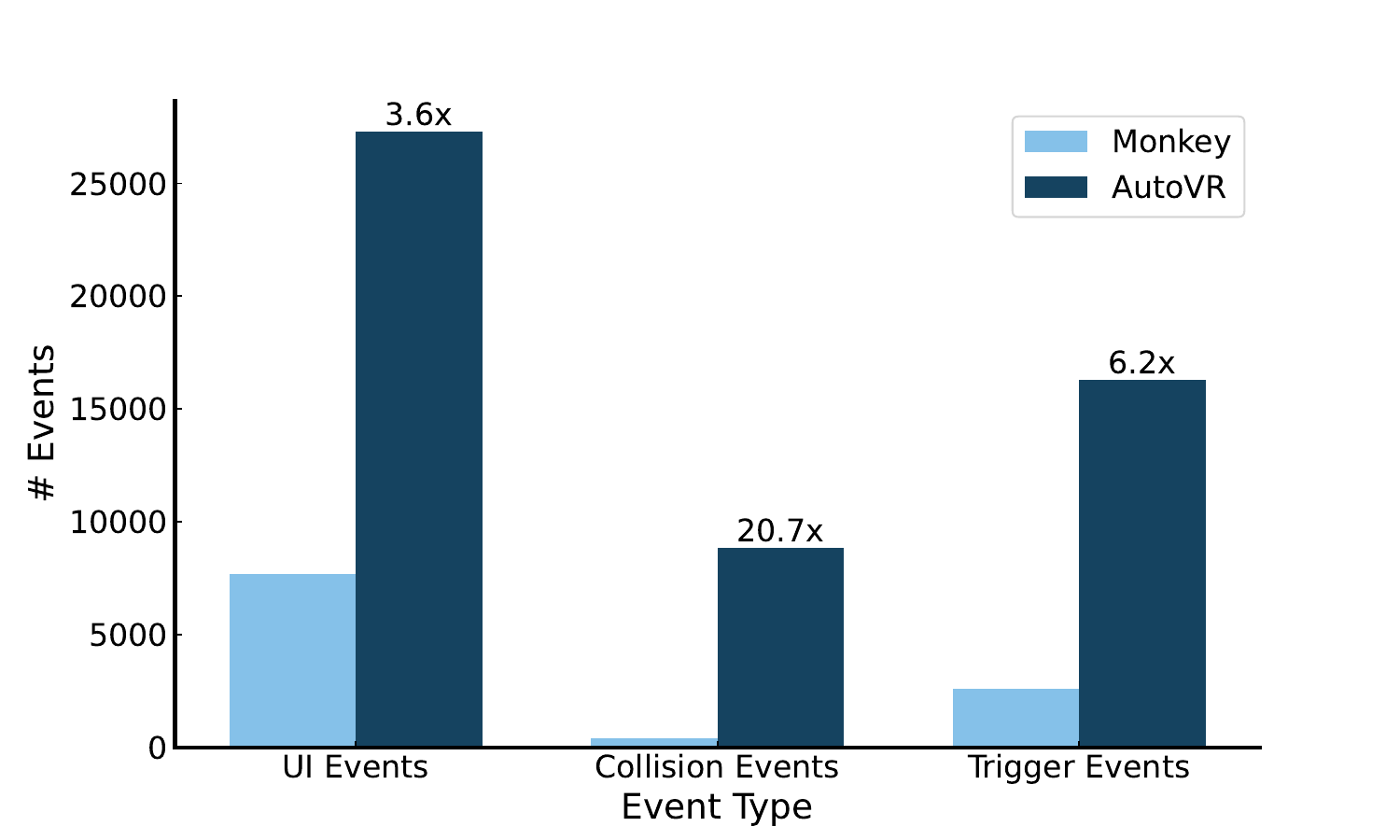}

      \vspace{-0.1in}
\caption{Comparison of the effectiveness (in terms of the number of triggering UI events and physics events) between \textsf{Monkey} and \sysname. }
      \vspace{-0.1in}
\label{side:performance}
\end{figure}
}

Furthermore, \sysname is context-aware, taking into account various game states such as scene iterations and the components of the scene (e.g., \texttt{GameObject}s). This enables it to trigger a greater number of events, thereby leading to the identification of more privacy exposures compared to \textsf{Monkey}. The comparative performance of \textsf{Monkey} and \sysname is additionally detailed in \autoref{tab:modelresults}. We notice that \sysname executes all events (i.e., UI events and physics events) far more frequently than \textsf{Monkey}. As such, a correlation can be determined by the number of privacy exposures detected by each tool versus the number of events executed, highlighting the advancements that \sysname has made over \textsf{Monkey}.

\subsection{Efficiency}
\label{sub:performance}

\begin{figure}[t]
    \centering
    \begin{subfigure}[t]{0.5\columnwidth}
        \centering
        \includegraphics[width=\columnwidth]{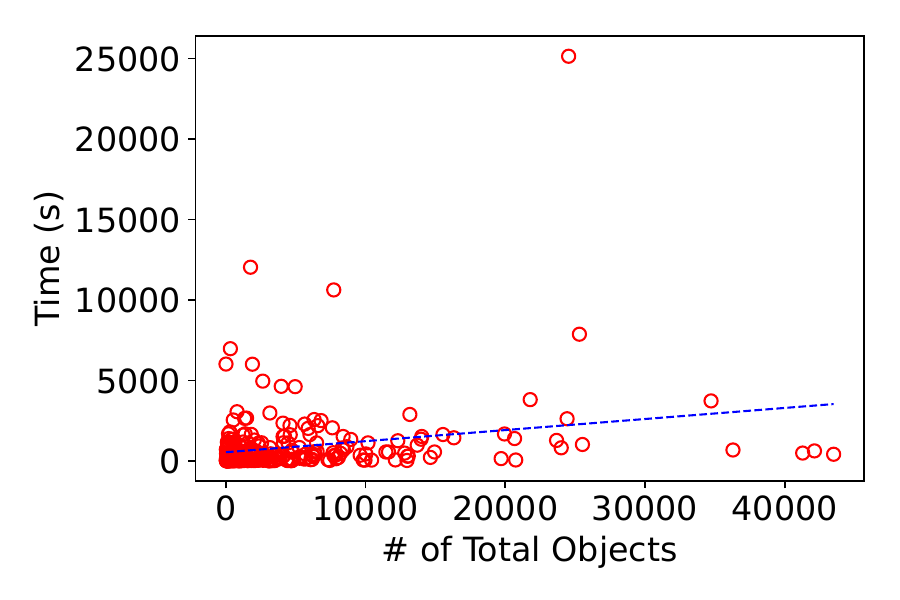}
        \caption{Objects vs Time}
        \label{time:objects_running}
    \end{subfigure}
    \hfill
    \begin{subfigure}[t]{0.49\columnwidth}
        \centering
        \includegraphics[width=\columnwidth]{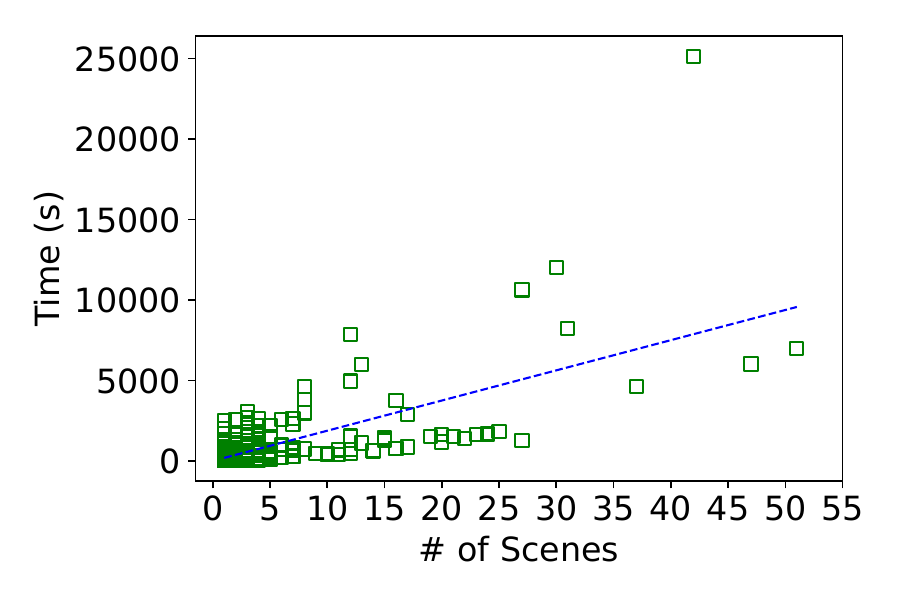}
        \caption{Scenes vs Time}
        \label{time:scenes_running}
    \end{subfigure}
    \vspace{-0.1in}
    \caption{Relationship of (a) total detected objects and (b) total scenes, to the total running time of \sysname for each VR app in our corpus, where (a) has a positive correlation coefficient of r = 0.24 and, (b) a stronger positive correlation coefficient r = 0.6721.}
    \vspace{-0.2in}
    \label{fig:object_scene_comparison}
\end{figure}

\color{black}

For the 366 VR apps, \sysname took a total of 81,398.92 seconds with an average running time of 222.40 seconds each. It is obvious that executing a random pre-computed set of events (\textsf{Monkey}) would have significantly shorter runtimes than a generative parsing and execution event model would (\sysname). 

From \autoref{fig:object_scene_comparison}, we notice a positive correlation between the number of \texttt{GameObject}s and runtime and an event stronger positive correlation between the number of scenes to the total runtime. While there is a clear correlation to the number of scenes and total objects to the running time, the event execution strategy also plays a factor. \sysname executes events in a DFS manner, and, for each newly found object, another DFS of the object's class fields is also performed for event identification. 
We show our calculation for the worst case running time in \autoref{eq:o}, where E = number of events, $P_c$ = collisionable objects, $P_t$ = triggerable objects, $C_t$ = class types, $C_f$ = fields per class type, $D$ = event dependencies. Because we are not memoizing ($C_t$ + $C_f$), after each event, \sysname must recalculate ($C_t$ + $C_f$) after each event invokes. As such, the running time scales higher depending on the combination of the three events.
\begin{equation}
O(E * (P_c + P_t + (C_t + C_f)) + D)
\label{eq:o}
\end{equation}
To quantify the efficiency in relation to the scene traversal events, EFC extraction, and event execution, we present \autoref{tab:benchmark}. We initially notice scene loading significantly contributes to the runtime, as \sysname waits 5 seconds for each time a scene loads, as such, the number scenes will scale with running time. Physics-based events are a significant contributor to the total running time of (2). For each physics event, all combinations of the available collisionable/triggerable \texttt{GameObject}s are executed with the EFC. In this case, $P_c$ + $P_t$ is larger than ($C_t$ + $C_f$), which is reflected in the stage's runtime. When bloating the number of UI events, the UI execution becomes the primary contributor to the total running time of (3). Notably, the UI execution from (3) starkly increases from (1)-(2). Because ($C_t$ + $C_f$) is recalculated after every UI event, the total ($C_t$ + $C_f$) will be significantly larger than both $P_c$ and $P_t$, resulting in the largest contributor to running time in (3).\looseness=-1

\begin{table}
\begin{minipage}{.5\linewidth}
\small
\centering
\begin{tabular}{l|c|c|c}
\toprule
& \multicolumn{3}{c}{Running Time (ms)} \\ \cline{1-4}
Stage Name & \textit{(1)} & \textit{(2)} & \textit{(3)} \\
\midrule
\textbf{Scene Loading} & 32,497 & 32,536 & 32,516 \\
\textbf{EFC Identification} & 859 & 1,280 & 1,196 \\
\textbf{UI Execution} & 9,074 & 9,034 & 46,833 \\
\textbf{Triggers Execution} & 21,064 & 86,220 & 21,093 \\
\textbf{Collisions Execution} & 22,129 & 91,072 & 22,151 \\
\bottomrule
\textbf{Miscellaneous} & 11,286 & 12,826 & 34,989 \\
\bottomrule
\textbf{Total} & 96,909 & 232,968 & 158,778 \\
\bottomrule
\end{tabular}
\end{minipage}
\caption{Relationship of time consuming \sysname stages to run time in milliseconds across three versions of the custom test app: (1) no changes, (2) with bloated physics events, (3) with bloated UI events. For (2) and (3), the bloated events are 3x from (1), including dependencies. }
\vspace{-0.2in}
\label{tab:benchmark}
\end{table}

\ignore{
\subsection{Effectiveness}
\label{sub:effectiveness}
\JK{Modify this once data is complete}

\ignore{
\begin{table*}[t]
\centering
\scriptsize
\setlength\extrarowheight{2pt} 
\begin{tabular}{c|c|c}\toprule
    \multicolumn{1}{l}{\bf Data Type} & \multicolumn{1}{|c}{\bf \# Triggered by \sysname} & \multicolumn{1}{|c}{\bf \# Triggered by \texttt{Monkey}} \\
    \midrule
    \multicolumn{1}{l}{\bf PII} \\
    \midrule
    \multirow{1}{*}{User ID} 
        & 2205 & 7 \\
    \multirow{1}{*}{Oculus ID} 
        & 2087 & 5 \\
    \multirow{1}{*}{Display Name} 
        & 2087 & 5 \\
    \multirow{1}{*}{User Profile Image} 
        & 2087 & 5 \\
    \multirow{1}{*}{Device Unique ID}
        & 3 & 2 \\
    \midrule
    \multicolumn{1}{l}{\bf Fingerprint} \\
    \midrule
    \multirow{1}{*}{Hardware Info} & 16 & 9 \\
    \midrule
    \multicolumn{1}{l}{\bf Communications} \\
    \midrule
    \multirow{1}{*}{Livestreaming} & 8 & 2 \\
    \multirow{1}{*}{User Friends} & 7 & 2 \\
    \midrule
    \multicolumn{1}{l}{\bf Biometrics} \\
    \midrule
    \multirow{1}{*}{Eye Tracking} & 1 & 0 \\
    \multirow{1}{*}{Microphone Usage} & 4 & 0\\    
    \midrule
    \multicolumn{1}{l}{\bf VR Data} \\
    \midrule
    \multirow{1}{*}{Play Boundary} & 10 & 10 \\
    \midrule
\end{tabular}
\caption{Privacy triggers invoked through \sysname compared to \texttt{Monkey}}
\label{tab:triggered}
\end{table*}
}
We have closely monitored \sysname's performance and metrics in 10 \texttt{Unity} VR games. We carefully selected such games that do not use APK signature verification to start the games. To measure the performance of \sysname, we compared the number of privacy triggers found using the Android \textsf{Monkey} tool, and with using \sysname. For \textsf{Monkey}, we set up an additional \texttt{frida-gadget} that injects our instrumentation code. This code contains all the private data trigger hooks and merely reports the triggers as \textsf{Monkey} progresses through the game. We summarize our findings \autoref{tab:triggered}. For each of the 10 games, we tested \textsf{Monkey} with the same number of EFCs found using \sysname. We can see that \sysname outperforms \textsf{Monkey} in triggering private data APIs. \textsf{Monkey} tries to invoke events by simulating physical touches that a normal Android app would take as input. As such, this conventional method of testing would underperform in an app that has no physical touch or press interactions. Because we are testing VR games, where inputs are more related towards controller cursor and movement events in a virtual environment, \textsf{Monkey} will not perform many promising EFCs like \sysname has.

However, we notice \textsf{Monkey} was still able to trigger many private data triggers, especially the user data. We notice most games will typically initialize all user data at the start of the game, without interaction from the user. As such, \textsf{Monkey} does not necessarily invoke these data triggers, but the game's initialization does. This goes the same way with \sysname. However, for \sysname, because we are able to traverse scenes (i.e., load and unload scenes of the game), we are triggering more of these initialization calls.

\subsection{Performance}

\JK{Modify this once data is complete}
\ignore {
\begin{figure}[t]
\centering
\scriptsize
\graphicspath{{./Figs/}}
\includegraphics[scale=.5, keepaspectratio]{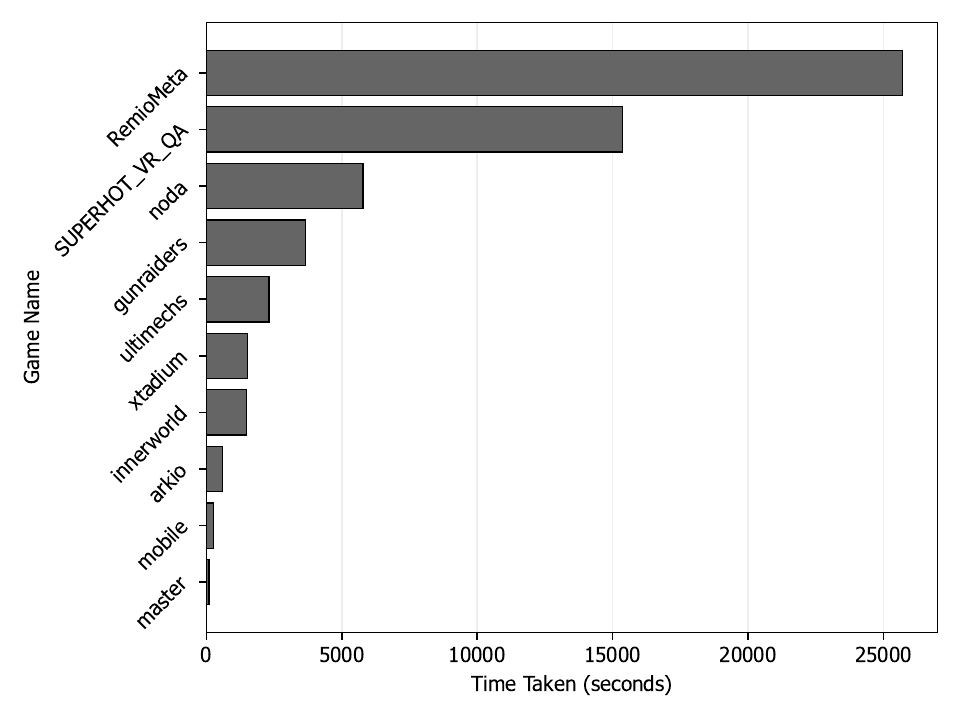}
\centering
\caption{Time taken for \sysname to analyze per game.}
\label{fig:times}
\end{figure}
}

\ignore {
\begin{table*}[t]
\centering
\setlength\tabcolsep{4pt}
\setlength\extrarowheight{2pt} 
\begin{tabular}{c||cc|cc|c||c}
\multirow{2}{*}{\textbf{Package Name}} & \multirow{2}{*}{\textbf{\# Scenes}} & \multirow{2}{*}{\textbf{\# Objects}} & \multicolumn{2}{c|}{\textbf{\# Events}} & \multirow{2}{*}{\textbf{\# Branches}} & \multirow{2}{*}{\textbf{Time Taken (seconds)}} \\
\cline{4-5}
 & & & \textbf{\# UnityEvents} & \textbf{\# Physics} & & \\ \hline
com.gpb.master & 3 & 1,299 & 46 & 0 & 90 & 95 \\
ob.opencampus.mobile & 9 & 4,142 & 202 & 0 & 305 & 261 \\
com.arkio.arkio & 1 & 7,799 & 800 & 0 & 1,179 & 602 \\
com.veryrealhelp.innerworld & 3 & 13,756 & 3,085 & 18 & 3,722 & 1,474 \\
com.ybvr.xtadium & 12 & 18,984 & 3,305 & 0 & 5,800 & 1,524 \\
com.resolutiongames.ultimechs & 12 & 216,419 & 9,398 & 0 & 9,764 & 2,315 \\
com.gunraiders & 8 & 224,076 & 2509 & 1,907 & 3,025 & 3,662 \\
com.codingleap.noda & 7 & 55,461 & 6,126 & 0 & 5,711 & 5,785 \\
unity.SUPERHOT\_Team.SUPERHOT\_VR\_QA & 18 & 211,390 & 336 & 16,827 & 554 & 15,367 \\
com.Remio.RemioMeta & 5 & 240,488 & 1,720 & 175,492 & 1,714 & 25,711 \\
\midrule
Total & 78 & 993,814 & 27,527 & 193,244 & 31,864 & 72,165
\end{tabular}
\caption{Collection statistics of \sysname for each game analyzed.}
\label{tab:stats}
\end{table*}


\begin{table*}[b]
\centering
\setlength\tabcolsep{4pt}
\setlength\extrarowheight{2pt} 
\begin{tabular}{c||c|cc|c||c}
\multirow{2}{*}{\textbf{Package Name}} & \multirow{2}{*}{\textbf{\# Objects}} & \multicolumn{2}{c|}{\textbf{\# Events Triggered}} & \multirow{2}{*}{\textbf{\# Branches}} & \multirow{2}{*}{\textbf{Time Taken (seconds)}} \\
\cline{3-4}
 & & \textbf{\# UnityEvents} & \textbf{\# Physics} & & \\ \hline
com.gpb.master &  1,299 & 31 & 0 & 90 & 95 \\
ob.opencampus.mobile & 4,142 & 303 & 0 & 305 & 261 \\
com.arkio.arkio & 7,799 & 2831 & 0 & 1,179 & 602 \\
com.veryrealhelp.innerworld & 13,756 & 1,027 & 18 & 3,722 & 1,474 \\
com.ybvr.xtadium  & 18,984 & 16,470 & 0 & 5,800 & 1,524 \\
com.resolutiongames.ultimechs & 216,419 & 4,255 & 0 & 9,764 & 2,315 \\
com.gunraiders & 224,076 & 2,727 & 1,907 & 3,025 & 3,662 \\
com.codingleap.noda & 55,461 & 13,311 & 0 & 5,711 & 5,785 \\
unity.SUPERHOT\_Team.SUPERHOT\_VR\_QA & 211,390 & 361 & 16,827 & 554 & 15,367 \\
com.Remio.RemioMeta & 240,488 & 8,755 & 175,492 & 1,714 & 25,711 \\
\midrule
Total & 993,814 & 50,071 & 193,244 & 31,864 & 72,165
\end{tabular}
\caption{Collection statistics of \sysname for each game analyzed.}
\label{tab:stats}
\end{table*}
}

In practice, fuzzers take a substantial amount of time to perform systematic testing. Fuzz testers are naturally time consuming as the focus of a fuzzer is to have a wide code coverage. As such, we want to compare and correlate time taken to the number of identified objects, events and collisions triggered, and branches resolved. We show these correlations in \autoref{tab:stats}. We further visualize the time taken per game in \autoref{fig:times}.

We notice that games with a large number of objects tend to increase the time taken to perform testing, however, this may not be the only case. We identified both \texttt{UnityEvents} and collisions as time-consuming processes. Firstly, for \texttt{UnityEvents}, we notice that more than one object may have the same EFC. As such, there are more EFCs triggered than there are EFCs. For each \texttt{UnityEvent} triggered, we also delay for 50 ms to avoid overloading the game. As such, the combination of finding, analyzing, and executing \texttt{UnityEvents} are one of the leading causes for time. Secondly, we notice games with many collision events tend to significantly increase the time taken. For example, we notice the game \textit{SUPERHOT} ~\cite{superhot}, a game where players simulate fights against enemy artificial intelligence, contained a significant number of collisions which caused the game to run for a substantial amount of time. We similarly see this for the game \texttt{RemioMeta} ~\cite{remio}. 

To simulate a collision, \sysname will attempt to move the position of a collidable game object to another collidable game object. Moving the position of game objects will take time as the game and system cannot process all the collisions at once without crashing. Furthermore, moving all objects instantaneously does not accurately represent the behavior of a player. As such, collision delays are needed. We notice that collision delay plays a big factor in the time taken. This is shown as \textit{RemioMeta} contained the most number of collisions, and thus the most amount of time taken.

We lastly identify that branch resolution correlates with some time taken, as branch exploration and dependency analysis naturally take a noticeable time to perform. We notice from \autoref{tab:stats} that games with more explored branches typically take more time. While this isn't necessarily the leading cause for longer times, we did notice some events will take a considerable time (i.e., 20-30 seconds) depending on the size and number of branches of the function.

}

\section{Discussion, Limitations, and Future Work}
\label{sec:future}

\vspace{-0.1in}

\bheading{Privacy Implications of Sensitive Data Flows}.
While \sysname was successfully able to extract sensitive data flows from VR apps, tracking the privacy implications requires knowledge of the context (e.g., purpose, notice and consent, law, etc.) and type (e.g., PII, fingerprinting, cookies, etc.) of data flows. This is because the data flows themselves do not determine a privacy concern, rather, a labeling of the outgoing data. For instance, as shown in \autoref{case:study}, the outgoing network packet finds an email and password, however, this data relates to the developer's PII data and not the user's. As the data flow collected is PII from the developer, this is not necessarily a privacy concern for the user. Additionally, the outgoing hostname is also relevant since first-party and third-party contexts are also needed. In \autoref{case:study2}, we notice the sensitive data flows are outgoing to \texttt{Unity} servers, a first-party context, not necessarily a privacy concern without the purpose of the data being collected. As such, additional analysis must be performed with context to identify the privacy implications, which is outside of the scope of \sysname. \looseness=-1

Nevertheless, to show that \sysname is still useful to help privacy concern analyses, we have collected the privacy policies from our corpus of VR apps and cross-compared them with the collected sensitive data flows using OVRSeen's improved PoliCheck~\cite{247632, 277092} tool. The results of which are shown in \autoref{tab:flowdata_consistency} in the Appendix for interested readers. Only 158 of our corpus of 366 VR apps contained privacy policies, and only 44 VR apps were able to be analyzed. We notice a significant number of results of \textit{vague} and \textit{omitted} network-to-consistency values from the privacy policies versus the traffic being collected. \textit{Omitted} (i.e., data type not found in the privacy policy) and \textit{vague} results (i.e., data collector is vague in the privacy policy), are an indication of leaking data flow, however, without the purpose of the data being collected, one cannot conclude that it is indeed a privacy concern.\looseness=-1

While the VR ecosystem continues to develop, there is still a lack of readily available privacy policies within both the Meta Quest and Side Quest app stores. In 2024, Guo et al.~\cite{guo2025empirical} has collected 900 VR apps from the SideQuest app store, and have found only 44.8\% contain privacy policies. Similarly, amongst our corpus of 366 VR apps, only 43.2\% contain privacy policies. Not only is there a considerable lack of privacy policies within the VR ecosystem, but also a significant number of inconsistent data collection policies found by OVRSeen~\cite{277092}. Yet, without these privacy policies, privacy analysis becomes significantly more challenging as this critical context is missing from the overall analysis.



\color{black}

\bheading{Comparison with state-of-the-art}. \sysname effectively exercises events within a VR game. Because these VR games are essentially APKs, it is only possible to exercise events using Android-supported event tools created before \sysname. \textsf{Monkey}, for example, is one such a tool. 
Similarly, existing state-aware UI exercising tools such as DroidBot~\cite{droidbot} and AutoDroid~\cite{autodroid} also utilize the same triggering functionality as \textsf{Monkey} by using MonkeyRunner~\cite{monkeyrunner} as the engine for triggering events. As MonkeyRunner triggers UI events by simulating screen-level human interactions (e.g., touch, swipes, double-taps), which are inconsistent event types within a VR environment, event invocation becomes much more difficult. Inherently, any Android-based event exercising tool are limited to 2D space, unlike VR, where events are expressed in a 3D space (e.g, event occurs \textbf{behind} the user's viewpoint will be unknown to traditional tools). Therefore, state-of-the-art tools such as DroidBot and AutoDroid consequently fail for VR apps.
\sysname, however, does not run into described problems. \sysname's automation relies on instructing events on the IL2CPP binary level, overcoming the screen image and VR controller dependency that current state-of-the-art tools run into. As such, \sysname significantly advances UI automation and dynamic analysis for VR games/apps.

\bheading{Integration of symbolic execution}.
\sysname provides the foundation of event identification and execution. This lays the groundwork for symbolic execution. Consequently, because the Unity Engine highly utilizes custom object types as the basis of the SDK, \texttt{UnityEvents} also support object types as inputs to execute events. However, identifying such inputs requires a deeper look into solving event constraints and is beyond the scope of this work. We foresee \sysname to expand and support solving. An SMT solver (e.g., the Z3 solver~\cite{z3prover})  will need to be integrated with EFCs that contain parameters that can be abstracted and taken as input for the SMT solver.
Encouragingly, \sysname event execution and state-modeling provides a first step towards understanding the dynamic paths of a game, opening up an opportunity for further investigation. \looseness=-1

\vspace{-0.1in}
\section{Related Work}
\label{sec:related}

\vspace{-0.1in}
\bheading{Privacy and Security in VR.}
Privacy risks in VR and head-mounted displays (HMDs) have been studied since the technology's inception~\cite{adams2018ethics,george2017seamless,vilk2015surroundweb}. Trimananda et al.~\cite{277092} audited network traffic to identify sensitive data risks, revealing that 70\% of Meta VR apps’ data flows were not disclosed to users. O'Brolcháin et al.~\cite{o2016convergence} explored ethical concerns, including privacy threats, social manipulation, and the blurred lines between real and virtual worlds. George et al.~\cite{george2017seamless} analyzed the usability and security of traditional authentication methods in VR, while Lou et al.~\cite{lou2019realistic} demonstrated real-time facial reconstruction through sensory input. \sysname complements these works by providing an automated framework to explore VR apps and expose privacy risks.\looseness=-1

\bheading{Android Testing.}
As Meta Quest runs on an Android OS variant, related works in Android UI testing are relevant. Gu et al.~\cite{gu2019practical} abstracted the Android GUI model to generate test cases, and Li et al.~\cite{droidbot} introduced a lightweight tool for dynamic, UI-guided test inputs. Su et al.~\cite{su2021fully} employed functional fuzz testing to uncover logic bugs, while Huang et al.~\cite{huang2019fuzzing} emphasized fuzzing Android apps in the context of data handling. More recently, Liu et al.~\cite{liu2023fill} leveraged Large Language Models (LLMs) for context-aware test input generation, and Ran et al.~\cite{ran2023badge} proposed prioritizing UI events to improve code coverage and detect unique crashes.\looseness=-1

State-of-the-art tools like \textsf{DroidBot}~\cite{droidbot} and \textsf{AutoDroid}~\cite{liu2023fill} rely on computer vision and tools such as \textsf{Minicap}~\cite{minicap} for state identification, but these methods are incompatible with Android 12, which powers Quest 2 devices. Additionally, they depend on \textsf{MonkeyRunner} for event triggering, which struggles with VR-specific applications. Despite advancements in Android testing, a critical gap remains in developing specialized dynamic tools for 3D engine-based apps and VR games on platforms like Meta Quest.\looseness=-1

\section{Conclusion}
\label{sec:conclude}

We have presented \sysname, a novel framework tailored for dynamically executing and exploring VR apps. By introspecting into the app's internal binary, \sysname overcomes the limitations of existing tools, efficiently identifying otherwise inaccessible events. Our empirical evaluation, compared against \textsf{Monkey}, illustrates \sysname's significant advancement in event exploration of VR environments by triggering events that lead to sensitive data exposures, thereby enhancing the security and privacy of VR apps. \looseness=-1

\section*{Acknowledgments}
\label{sec:acknowledgments}
We would like to thank the anonymous reviewers and our shepherd for providing their feedback that improved this paper. Additionally, we would like to give special thanks to Jerry Liang and Edward Liu for providing support for this work. This work was supported by a research grant from \textit{Meta}, in addition to providing a test account and an unlocked Meta Quest 2 device.


\noindent
\section*{Ethics Considerations}
We ensured that ethics were considered early in the experimentation process with AutoVR. To collect user data and instrumentation, three locked Meta Quest 2 devices were purchased and used with test accounts. Additionally, a fourth unlocked (i.e., rootable) Quest 2 device was provided by \textit{Meta}, along with a test account that had entitlements to the tested 366 VR apps used in the evaluation (\S\ref{sec:evaluation}). \textit{Meta} has given explicit permission to instrument on VR apps from the Meta Quest store, usage, and instrumentation of AutoVR on the rooted device. Therefore, AutoVR addresses the following concerns: (a) experimented VR apps were entitled to the instrumenting Meta Quest account, and (b) instrumented on devices with owner permission. However, we must note that any misuse of AutoVR may occur if (a) and (b) are violated. Additionally, the data collected from the VR devices and subsequently the test accounts will be stripped of potentially identifiable information (PII) leaked by the developer of each VR app, prior to the release of the network data.

\section*{Open Science}
The source code of AutoVR is available on GitHub under the MIT license. Additionally, raw data collected from the AutoVR experiments depicted in the evaluation (\S\ref{sec:evaluation}), along with the stripped network data collected by \textit{AntMonitor}. These artifacts can be found at \url{https://doi.org/10.5281/zenodo.15636793}.



\balance

\bibliographystyle{plain}
\bibliography{paper}

\begin{thebibliography}{10}

\bibitem{Button1}
{B}utton | {U}nity {U}{I} | 1.0.0 --- 1.0.
\newblock \url{https://docs.unity3d.com/Packages/com.unity.ugui@1.0/manual/script-Button.html}.
\newblock [Accessed 03-06-2024].

\bibitem{epicgamesDrivingUpdates}
{D}riving {U}{I} {U}pdates with {E}vents in {U}nreal {E}ngine | {U}nreal {E}ngine 5.5 {D}ocumentation | {E}pic {D}eveloper {C}ommunity --- dev.epicgames.com.
\newblock \url{https://dev.epicgames.com/documentation/en-us/unreal-engine/driving-ui-updates-with-events-in-unreal-engine}.
\newblock [Accessed 27-05-2025].

\bibitem{frida}
{frida - a world-class dynamic instrumentation toolkit}.
\newblock \url{https://frida.re/}.

\bibitem{interceptor}
Frida interceptor.
\newblock \url{https://frida.re/docs/javascript-api/#interceptor}, journal={Frida}, year={2023}, month={Mar}.

\bibitem{il2cpp}
Il2cpp overview.
\newblock \url{https://docs.unity3d.com/Manual/IL2CPP.html}.

\bibitem{epicgamesLevelsUnreal}
{L}evels in {U}nreal {E}ngine | {U}nreal {E}ngine 5.5 {D}ocumentation | {E}pic {D}eveloper {C}ommunity --- dev.epicgames.com.
\newblock \url{https://dev.epicgames.com/documentation/en-us/unreal-engine/levels-in-unreal-engine}.
\newblock [Accessed 27-05-2025].

\bibitem{monkeyrunner}
monkeyrunner | android studio | android developers.
\newblock \url{https://developer.android.com/studio/test/monkeyrunner}.

\bibitem{mono}
Mono overview.
\newblock \url{https://docs.unity3d.com/Manual/Mono.html}, journal={Unity}, author={Technologies, Unity}.

\bibitem{mike_2020}
Oculus quest 2 game development options.
\newblock \url{https://gamefromscratch.com/oculus-quest-2-game-development-options/}, journal={GameFromScratch.com}, author={Mike}, year={2020}, month={Oct}.

\bibitem{il2cppdumper}
Perfare/il2cppdumper: Unity il2cpp reverse engineer.
\newblock \url{https://github.com/Perfare/Il2CppDumper}, journal={GitHub}, author={Perfare}.

\bibitem{UnityLau99:online}
Unity launches beta program for visionos — enabling unity developers to create games and apps for apple vision pro | business wire.
\newblock \url{https://www.businesswire.com/news/home/20230719202814/en/Unity-Launches-Beta-Program-for-visionOS-%E2%80%94-Enabling-Unity-Developers-to-Create-Games-and-Apps-for-Apple-Vision-Pro}.
\newblock (Accessed on 07/31/2023).

\bibitem{unity}
Unity real-time development platform.
\newblock \url{https://unity.com/}.

\bibitem{VirtualR15:online}
Virtual reality (vr) market size, share, growth \& trends.

\bibitem{katycode}
Il2cpp reverse engineering part 1: Hello world and the il2cpp toolchain.
\newblock \url{https://katyscode.wordpress.com/2020/06/24/il2cpp-part-1/}, Dec 2020.

\bibitem{monkey}
{ UI/Application Exerciser Monkey}.
\newblock \url{https://developer.android.com/studio/test/other-testing-tools/monkey}, Jan 2022.
\newblock [Online; accessed 1. Mar. 2023].

\bibitem{gadget}
Gadget.
\newblock \url{https://frida.re/docs/gadget/}, Mar 2023.

\bibitem{unityui}
{The interaction components in the UI system handle interaction, such as mouse or touch events and interaction using a keyboard or controller.}
\newblock \url{https://docs.unity3d.com/Packages/com.unity.ugui@1.0/manual/comp-UIInteraction.html}, Jun 2023.
\newblock [Online; accessed 15. Jul. 2023].

\bibitem{adams2018ethics}
Devon Adams, Alseny Bah, Catherine Barwulor, Nureli Musaby, Kadeem Pitkin, and Elissa~M Redmiles.
\newblock Ethics emerging: the story of privacy and security perceptions in virtual reality.
\newblock In {\em Fourteenth Symposium on Usable Privacy and Security (SOUPS 2018)}, pages 427--442, 2018.

\bibitem{247632}
Benjamin Andow, Samin~Yaseer Mahmud, Justin Whitaker, William Enck, Bradley Reaves, Kapil Singh, and Serge Egelman.
\newblock Actions speak louder than words: {Entity-Sensitive} privacy policy and data flow analysis with {PoliCheck}.
\newblock In {\em 29th USENIX Security Symposium (USENIX Security 20)}, pages 985--1002. USENIX Association, August 2020.

\bibitem{Antonishyn2020}
Mykhailo Antonishyn.
\newblock Four ways to bypass android {SSL}. verification and certificate pinning.
\newblock {\em Transfer of Innovative Technologies}, 3(1):96--99, September 2020.

\bibitem{bacis2024assessing}
Enrico Bacis, Igor Bilogrevic, Robert Busa-Fekete, Asanka Herath, Antonio Sartori, and Umar Syed.
\newblock Assessing web fingerprinting risk.
\newblock In {\em Companion Proceedings of the ACM on Web Conference 2024}, pages 245--254, 2024.

\bibitem{billwagner}
BillWagner.
\newblock Attributes and reflection.
\newblock \url{https://learn.microsoft.com/en-us/dotnet/csharp/advanced-topics/reflection-and-attributes/}.

\bibitem{george2017seamless}
Ceenu George, Mohamed Khamis, Emanuel von Zezschwitz, Marinus Burger, Henri Schmidt, Florian Alt, and Heinrich Hussmann.
\newblock Seamless and secure vr: Adapting and evaluating established authentication systems for virtual reality.
\newblock NDSS, 2017.

\bibitem{gu2019practical}
Tianxiao Gu, Chengnian Sun, Xiaoxing Ma, Chun Cao, Chang Xu, Yuan Yao, Qirun Zhang, Jian Lu, and Zhendong Su.
\newblock Practical gui testing of android applications via model abstraction and refinement.
\newblock In {\em 2019 IEEE/ACM 41st International Conference on Software Engineering (ICSE)}, pages 269--280. IEEE, 2019.

\bibitem{guo2025empirical}
Hanyang Guo, Hong-Ning Dai, Xiapu Luo, Gengyang Xu, Fengliang He, and Zibin Zheng.
\newblock An empirical study on meta virtual reality applications: Security and privacy perspectives.
\newblock {\em IEEE Transactions on Software Engineering}, 2025.

\bibitem{Hauwert_2014}
Ralph Hauwert.
\newblock The future of scripting in unity, May 2014.

\bibitem{huang2019fuzzing}
Xinyue Huang, Anmin Zhou, Peng Jia, Luping Liu, and Liang Liu.
\newblock Fuzzing the android applications with http/https network data.
\newblock {\em IEEE Access}, 7:59951--59962, 2019.

\bibitem{Le2015}
Anh Le, Janus Varmarken, Simon Langhoff, Anastasia Shuba, Minas Gjoka, and Athina Markopoulou.
\newblock Antmonitor: A system for monitoring from mobile devices.
\newblock In {\em Proceedings of the 2015 ACM SIGCOMM Workshop on Crowdsourcing and Crowdsharing of Big (Internet) Data}, pages 15--20, 2015.

\bibitem{droidbot}
Yuanchun Li, Ziyue Yang, Yao Guo, and Xiangqun Chen.
\newblock Droidbot: a lightweight ui-guided test input generator for android.
\newblock In {\em 2017 IEEE/ACM 39th International Conference on Software Engineering Companion (ICSE-C)}, pages 23--26, 2017.

\bibitem{liu2023fill}
Zhe Liu, Chunyang Chen, Junjie Wang, Xing Che, Yuekai Huang, Jun Hu, and Qing Wang.
\newblock Fill in the blank: Context-aware automated text input generation for mobile gui testing.
\newblock In {\em 2023 IEEE/ACM 45th International Conference on Software Engineering (ICSE)}, pages 1355--1367. IEEE, 2023.

\bibitem{lou2019realistic}
Jianwen Lou, Yiming Wang, Charles Nduka, Mahyar Hamedi, Ifigeneia Mavridou, Fei-Yue Wang, and Hui Yu.
\newblock Realistic facial expression reconstruction for vr hmd users.
\newblock {\em IEEE Transactions on Multimedia}, 22(3):730--743, 2019.

\bibitem{ovrstore}
Meta.
\newblock Oculus quest store: Vr games, apps, \&; more.
\newblock \url{https://www.oculus.com/experiences/quest/}.

\bibitem{moody_2021}
Meaghan Moody.
\newblock Oculus quest 2 specifications.
\newblock \url{https://studiox.lib.rochester.edu/oculus-quest-2-specifications/}, Sep 2021.

\bibitem{minicap}
OpenSTF.
\newblock Openstf/minicap: Stream real-time screen capture data out of android devices.

\bibitem{o2016convergence}
Fiachra O’Brolch{\'a}in, Tim Jacquemard, David Monaghan, Noel O’Connor, Peter Novitzky, and Bert Gordijn.
\newblock The convergence of virtual reality and social networks: threats to privacy and autonomy.
\newblock {\em Science and engineering ethics}, 22:1--29, 2016.

\bibitem{ran2023badge}
Dezhi Ran, Hao Wang, Wenyu Wang, and Tao Xie.
\newblock Badge: Prioritizing ui events with hierarchical multi-armed bandits for automated ui testing.
\newblock In {\em 2023 IEEE/ACM 45th International Conference on Software Engineering (ICSE)}, pages 894--905. IEEE, 2023.

\bibitem{rzig2023virtual}
Dhia~Elhaq Rzig, Nafees Iqbal, Isabella Attisano, Xue Qin, and Foyzul Hassan.
\newblock Virtual reality (vr) automated testing in the wild: A case study on unity-based vr applications.
\newblock In {\em Proceedings of the 32nd ACM SIGSOFT International Symposium on Software Testing and Analysis}, pages 1269--1281, 2023.

\bibitem{sensepost}
Sensepost.
\newblock objection - runtime mobile exploration.
\newblock \url{https://github.com/sensepost/objection}.

\bibitem{su2021fully}
Ting Su, Yichen Yan, Jue Wang, Jingling Sun, Yiheng Xiong, Geguang Pu, Ke~Wang, and Zhendong Su.
\newblock Fully automated functional fuzzing of android apps for detecting non-crashing logic bugs.
\newblock {\em Proceedings of the ACM on Programming Languages}, 5(OOPSLA):1--31, 2021.

\bibitem{events}
Unity Technologies.
\newblock {U}nity - {M}anual: {E}vent {F}unctions --- docs.unity3d.com.
\newblock \url{https://docs.unity3d.com/Manual/EventFunctions.html}.
\newblock [Accessed 03-06-2024].

\bibitem{gameobject1}
Unity Technologies.
\newblock {U}nity - {M}anual: {G}ame{O}bject --- docs.unity3d.com.
\newblock \url{https://docs.unity3d.com/Manual/class-GameObject.html}.
\newblock [Accessed 03-06-2024].

\bibitem{rulematrix}
Unity Technologies.
\newblock {U}nity - {M}anual: {I}nteraction between collider types --- docs.unity3d.com.
\newblock \url{https://docs.unity3d.com/Manual/collider-types-interaction.html}.
\newblock [Accessed 02-06-2024].

\bibitem{colliders}
Unity Technologies.
\newblock {U}nity - {M}anual: {I}ntroduction to collider types --- docs.unity3d.com.
\newblock \url{https://docs.unity3d.com/Manual/collider-types-introduction.html}.
\newblock [Accessed 03-06-2024].

\bibitem{unityexecutionorder}
Unity Technologies.
\newblock {U}nity - {M}anual: {O}rder of execution for event functions --- docs.unity3d.com.
\newblock \url{https://docs.unity3d.com/2021.3/Documentation/Manual/ExecutionOrder.html}.
\newblock [Accessed 27-05-2025].

\bibitem{Button2}
Unity Technologies.
\newblock {U}nity - {S}cripting {A}{P}{I}: {B}utton --- docs.unity3d.com.
\newblock \url{https://docs.unity3d.com/2018.2/Documentation/ScriptReference/UI.Button.html}.
\newblock [Accessed 03-06-2024].

\bibitem{fixedupdate}
Unity Technologies.
\newblock {U}nity - {S}cripting {A}{P}{I}: {M}ono{B}ehaviour.{F}ixed{U}pdate() --- docs.unity3d.com.
\newblock \url{https://docs.unity3d.com/2022.3/Documentation/ScriptReference/MonoBehaviour.FixedUpdate.html}.
\newblock [Accessed 27-05-2025].

\bibitem{ieventsystemhandler}
Unity Technologies.
\newblock {U}nity - {S}cripting {A}{P}{I}: {I}{E}vent{S}ystem{H}andler --- docs.unity3d.com.
\newblock \url{https://docs.unity3d.com/2019.1/Documentation/ScriptReference/EventSystems.IEventSystemHandler.html}, August 2019.
\newblock [Accessed 14-10-2023].

\bibitem{277092}
Rahmadi Trimananda, Hieu Le, Hao Cui, Janice~Tran Ho, Anastasia Shuba, and Athina Markopoulou.
\newblock {OVRseen}: Auditing network traffic and privacy policies in oculus {VR}.
\newblock In {\em 31st USENIX Security Symposium (USENIX Security 22)}, pages 3789--3806, Boston, MA, August 2022. USENIX Association.

\bibitem{collider}
Unity.
\newblock Collider.
\newblock \url{https://docs.unity3d.com/ScriptReference/Collider.html}.

\bibitem{component}
Unity.
\newblock Component.
\newblock \url{https://docs.unity3d.com/ScriptReference/Component.html}.

\bibitem{gameobject}
Unity.
\newblock Gameobject.
\newblock \url{https://docs.unity3d.com/ScriptReference/GameObject.html}.

\bibitem{stripping}
Unity.
\newblock Managed code stripping.
\newblock \url{https://docs.unity3d.com/Manual/ManagedCodeStripping.html}.

\bibitem{scenemanager}
Unity.
\newblock Scenemanager.
\newblock \url{https://docs.unity3d.com/ScriptReference/SceneManagement.SceneManager.html}.

\bibitem{vilk2015surroundweb}
John Vilk, David Molnar, Benjamin Livshits, Eyal Ofek, Chris Rossbach, Alexander Moshchuk, Helen~J Wang, and Ran Gal.
\newblock Surroundweb: Mitigating privacy concerns in a 3d web browser.
\newblock In {\em 2015 IEEE Symposium on Security and Privacy}, pages 431--446. IEEE, 2015.

\bibitem{autodroid}
Hao Wen, Yuanchun Li, Guohong Liu, Shanhui Zhao, Tao Yu, Toby Jia-Jun Li, Shiqi Jiang, Yunhao Liu, Yaqin Zhang, and Yunxin Liu.
\newblock Autodroid: Llm-powered task automation in android, 2024.

\bibitem{z3prover}
Z3Prover.
\newblock Z3prover/z3: The z3 theorem prover.
\newblock \url{https://github.com/Z3Prover/z3}.

\bibitem{279950}
Chaoshun Zuo and Zhiqiang Lin.
\newblock Playing without paying: Detecting vulnerable payment verification in native binaries of unity mobile games.
\newblock In {\em 31st USENIX Security Symposium (USENIX Security 22)}, pages 3093--3110, Boston, MA, August 2022. USENIX Association.

\end{thebibliography}
\appendix
\newpage


\ignore{\begin{table*}[t]
\scriptsize
\centering
\begin{tabular}{c|c|c|c}\toprule
    \multicolumn{1}{c}{\bf API Name} & \multicolumn{1}{|c}{\bf Return Type} & \multicolumn{1}{|c}{\bf Argument Types} & \multicolumn{1}{|c}{\bf Description}  \\
    \midrule
    \hline
    \multicolumn{4}{c}{\bf Classes} \\
    \bottomrule
    \multicolumn{1}{l}{il2cpp\_class\_get\_field\_from\_name} & \multicolumn{1}{l}{field pointer} & \multicolumn{1}{l}{class pointer, string} & \multicolumn{1}{l}{Get class field handle from name.} \\ \hline
    \multicolumn{1}{l}{il2cpp\_class\_get\_fields} & \multicolumn{1}{l}{fields array pointer} & \multicolumn{1}{l}{class pointer, class type} & \multicolumn{1}{l}{Get field handles from class.} \\ \hline
    \multicolumn{1}{l}{il2cpp\_class\_get\_method\_from\_name} & \multicolumn{1}{l}{method pointer} & \multicolumn{1}{l}{class pointer, string, int} & \multicolumn{1}{l}{Get class method handle from method name and parameter count.} \\ \hline
    \multicolumn{1}{l}{il2cpp\_class\_get\_methods} & \multicolumn{1}{l}{methods array pointer} & \multicolumn{1}{l}{class pointer, pointer, class type} & \multicolumn{1}{l}{Get all method handles from class.} \\ \hline
    \multicolumn{1}{l}{il2cpp\_class\_get\_name} & \multicolumn{1}{l}{string} & \multicolumn{1}{l}{class pointer} & \multicolumn{1}{l}{Get name from class handle.} \\ \hline
    \multicolumn{1}{l}{il2cpp\_class\_get\_namespace} & \multicolumn{1}{l}{string} & \multicolumn{1}{l}{class pointer} & \multicolumn{1}{l}{Get class namespace name from class handle.} \\ \hline
    \multicolumn{1}{l}{il2cpp\_class\_get\_parent} & \multicolumn{1}{l}{parent class pointer} & \multicolumn{1}{l}{class pointer} & \multicolumn{1}{l}{Get class' parent handle.} \\ \hline
    \multicolumn{1}{l}{il2cpp\_class\_get\_type} & \multicolumn{1}{l}{class type pointer} & \multicolumn{1}{l}{class pointer} & \multicolumn{1}{l}{Get class type from class handle.} \\ \hline
    \multicolumn{1}{l}{il2cpp\_class\_is\_assignable\_from} & \multicolumn{1}{l}{bool} & \multicolumn{1}{l}{class pointer, other class pointer} & \multicolumn{1}{l}{Test if class is assignable from another class.} \\ \hline
    \bottomrule
    \multicolumn{4}{c}{\bf Fields} \\ 
    \bottomrule
    \multicolumn{1}{l}{il2cpp\_field\_get\_parent} & \multicolumn{1}{l}{parent class pointer} & \multicolumn{1}{l}{class pointer} & \multicolumn{1}{l}{Get parent class of field handle.} \\ \hline
    \multicolumn{1}{l}{il2cpp\_field\_get\_name} & \multicolumn{1}{l}{field pointer} & \multicolumn{1}{l}{string} & \multicolumn{1}{l}{Get field name from field handle.} \\ \hline
    \multicolumn{1}{l}{il2cpp\_field\_get\_offset} & \multicolumn{1}{l}{int32} & \multicolumn{1}{l}{field pointer} & \multicolumn{1}{l}{Get class field offset from field handle.} \\ \hline
    \multicolumn{1}{l}{il2cpp\_field\_static\_get\_value} & \multicolumn{1}{l}{void} & \multicolumn{1}{l}{field pointer, value pointer} & \multicolumn{1}{l}{Get field handle's static value into value pointer.} \\ \hline
    \multicolumn{1}{l}{il2cpp\_field\_get\_type} & \multicolumn{1}{l}{class type pointer} & \multicolumn{1}{l}{field pointer} & \multicolumn{1}{l}{Get type of field from field handle.} \\ \hline
    \multicolumn{1}{l}{il2cpp\_field\_is\_static} & \multicolumn{1}{l}{bool} & \multicolumn{1}{l}{field pointer} & \multicolumn{1}{l}{Test if field is static from field handle.} \\ \hline
    \bottomrule
    \multicolumn{4}{c}{\bf Methods} \\
    \bottomrule
    \multicolumn{1}{l}{il2cpp\_method\_get\_class} & \multicolumn{1}{l}{class pointer} & \multicolumn{1}{l}{method pointer} & \multicolumn{1}{l}{Get residing class handle from method handle.} \\ \hline
    \multicolumn{1}{l}{il2cpp\_method\_get\_name} & \multicolumn{1}{l}{string} & \multicolumn{1}{l}{method pointer} & \multicolumn{1}{l}{Get method name from method handle.} \\ \hline
    \multicolumn{1}{l}{il2cpp\_method\_get\_object} & \multicolumn{1}{l}{object pointer} & \multicolumn{1}{l}{method pointer} & \multicolumn{1}{l}{Get object handle from method handle.} \\ \hline
    \multicolumn{1}{l}{il2cpp\_method\_get\_param\_count} & \multicolumn{1}{l}{uint8} & \multicolumn{1}{l}{method pointer} & \multicolumn{1}{l}{Get number of parameters from method.} \\ \hline
    \multicolumn{1}{l}{il2cpp\_method\_get\_parameter\_name} & \multicolumn{1}{l}{parameters pointer} & \multicolumn{1}{l}{string, uint32} & \multicolumn{1}{l}{Get parameter handle from name and parameter index.} \\ \hline
    \multicolumn{1}{l}{il2cpp\_method\_get\_pointer} & \multicolumn{1}{l}{virtual address pointer} & \multicolumn{1}{l}{method pointer} & \multicolumn{1}{l}{Get virtual address from method handle.} \\ \hline
    \multicolumn{1}{l}{il2cpp\_method\_get\_return\_type} & \multicolumn{1}{l}{class type pointer} & \multicolumn{1}{l}{method pointer} & \multicolumn{1}{l}{Get return type of method.} \\ \hline
    \bottomrule
    \multicolumn{4}{c}{\bf Objects} \\
    \bottomrule
    \multicolumn{1}{l}{il2cpp\_object\_get\_class} & \multicolumn{1}{l}{object pointer} & \multicolumn{1}{l}{class pointer} & \multicolumn{1}{l}{Get class handle from object.} \\ \hline
    \multicolumn{1}{l}{il2cpp\_object\_new} & \multicolumn{1}{l}{object pointer} & \multicolumn{1}{l}{class pointer} & \multicolumn{1}{l}{Create a new object from class handle.} \\ \hline
    \multicolumn{1}{l}{il2cpp\_capture\_memory\_snapshot} & \multicolumn{1}{l}{objects array pointer} & \multicolumn{1}{l}{void} & \multicolumn{1}{l}{Creates a memory snapshot of all GC handles.} \\ \hline
    \bottomrule
\end{tabular}
\caption{IL2CPP Runtime Library API}
\label{tab:il2cppapi}
\end{table*}



}

\ignore{
\begin{table*}[htbp]
\centering
\scriptsize
\setlength\extrarowheight{2pt} 
    \begin{tabular}{|rllcc|ccccc|c|}
    \hline
    \multicolumn{5}{|c|}{\textbf{Metadata}} & \multicolumn{6}{c|}{\textbf{Performance}} \\
    \hline
    \textbf{App Package Name} & \textbf{Version} & \textbf{Categories} & \textbf{\# Downloads} & \textbf{Size (MB)} & \textbf{\# Scenes} & \textbf{\# GameObjects} & \textbf{\# UI events} & \textbf{\# Triggers} & \textbf{\# Collisions} & \textbf{Time (min)} \\
    \hline
    com.PBVR.Playground & v1.3.4 & fps   & 740,254 & 352.3 & 1     & 5,731 & 6     & 0     & 0     & 104.7 \\
    com.noowanda.discoverylite & 2.7.2.1 Lite & other & 549,749 & 70.9  & 1     & 580   & 1     & 0     & 0     & 17.1 \\
    com.FnafSLVR.SLVR & 1.0.1 & horror & 374,320 & 434.6 & 12    & 1,783 & 22    & 0     & 0     & 405.1 \\
    com.PavelMarceluch.VRtuos & 2.0.9b & education & 227,632 & 135.7 & 3     & 245   & 10    & 0     & 0     & 134.2 \\
    com.EngineOrganic.HAX & v0.161 & fps   & 188,513 & 433.6 & 5     & 10,871 & 192   & 22     & 390     & 343.32 \\
    com.FlodLab.OceanCraft & v0.0.10 & survival & 175,255 & 161.5 & 2     & 1,705 & 10    & 0     & 0     & 84.56 \\
    com.revomon.vr & v2.4.0 & adventure & 143,276 & 93.9  & 1     & 277   & 1     & 0     & 0     & 9.9 \\
    com.swearl.playa & 2.1.6 & video player & 133,840 & 159.8 & 6     & 138   & 15    & 0     & 0     & 192.1 \\
    com.Sol5Studios.TheSilkworm & release & climbing & 127,431 & 256.3 & 2     & 303   & 3     & 0     & 0     & 41.4 \\
    com.Virtuleap.Enhance & 2.6.1.0 & education & 88,571 & 135.7 & 17    & 1,783 & 0     & 0     & 0     & 1354.7 \\
    \hline
    \end{tabular}%
     \caption{The results of the performance data for 10 random VR apps. \yj{What does time mean? What is the unit of time? For static analysis or dynamic testing? Aggregated statsitcis}\ZQ{This table has to be explained. The main body should also refer to this table. Basically, to show the details/}}
  \label{tab:performance}%
\end{table*}%

}

\ignore{

\begin{table}[]
\centering
\scriptsize
\setlength\extrarowheight{1pt} 
\begin{tabular}{|l|r|r|r|r|r|}
\hline

\multirow{2}{*}{\textbf{Sensitive Data Type}} & \multicolumn{5}{c|}{\textbf{\# Leaks}}                                                                                                                            \\ \cline{2-6} 
                                   & \multicolumn{1}{l|}{\textbf{AppA}} & \multicolumn{1}{l|}{\textbf{AppB}} & \multicolumn{1}{l|}{\textbf{AppC}} & \multicolumn{1}{l|}{\textbf{AppD}} & \textbf{AppE} \\ \hline\hline
                                   
ANALYTIC&14&22&20&12&13\\
APP\_INFO&141&102&123&70&79\\
BUILD\_INFO&57&28&49&23&37\\
COOKIE&0&0&416&0&0\\
CPU\_INFO&60&36&12&13&35\\
DEVICE\_ID&27&46&25&24&5\\
DEVICE\_INFO&260&108&112&94&217\\
EMAIL&0&0&0&0&0\\
EVENT\_INFO&14&22&20&12&25\\
GPU\_INFO&320&36&180&78&259\\
INSTALL\_INFO&60&36&24&26&51\\
JAIL&30&18&12&13&16\\
LANGUAGE&40&0&24&8&52\\
OPENGL\_VERSION&30&18&12&13&35\\
OS\_INFO&20&0&12&4&16\\
PLATFORM\_INFO&134&92&98&64&94\\
PLAY\_SESSION&17&28&25&15&6\\
POSITION&1&0&0&0&0\\
SCREEN\_INFO&90&18&48&25&64\\
SCRIPTING\_BACKEND&20&0&12&4&16\\
SENSOR\_DATA&20&0&12&4&16\\
SESSION\_DATA&54&56&62&34&26\\
TRACKING&3&0&0&0&0\\
UNITY\_VERSION&183&112&1,423&80&249\\
USER\_ID&37&28&37&19&21\\
VR\_FIELD\_OF\_VIEW&10&18&0&9&0\\
VR\_PLAY\_AREA&10&18&0&9&0\\
VR\_USER\_DEVICE\_IPD&10&18&0&9&0\\ \hline
\end{tabular}
\caption{The top 5 games that leak the most sensitive information. Note that the specific name of these apps: \\
AppA: com.megames.deadscope \\ AppB: com.aura.Arrows \\ AppC: com.resolutiongames.baitsantacruz \\ AppD: com.PBVR.Playground\\ AppE: com.veryrealhelp.innerworld
\ZQ{Move this to appendix}
}
\end{table}
}

\section{Case Study} While perusing the collected data exposures for \sysname's traffic, we noticed incoming and outgoing network requests between the VR device and external servers. Notably, the one app that exposes \texttt{EMAIL} was identified traffic coming from \texttt{ob.opencampus.mobile}, which contained alarming HTTP packets that include the plaintext string of an individual's email and password. Specifically, as shown in \autoref{case:study}, we notice that the endpoint to the incoming and outgoing traffic is linked to a \texttt{gcsvrapi.worldbank.org} domain. However, when attempting to invoke the same exposure behavior with \textsf{Monkey}, none of this traffic was identified. This is also supported by \autoref{fig:comparemonkey}. This highlights the significance of \sysname's ability to uncover events at the binary level, invoking events accurately and triggering potentially damaging privacy exposures. \looseness=-1

\begin{figure}
\begin{scriptsize}
\begin{verbatim}
ob.opencampus.mobile#0.5P???
??email=********%40protonmail.com&password=******
%21ob.opencampus.mobile#0.5?P??
?POST /api/login HTTP/1.1
Host: gcsvrapi.worldbank.org
Accept-Encoding: gzip, identity
Connection: Keep-Alive, TE
TE: identity
User-Agent: BestHTTP 1.12.1
Content-Type: application/x-www-form-urlencoded
Content-Length: 54

ob.opencampus.mobile#0.5P????ebbEb??????P??IB
GET /api/featured?unity=true&wm=false HTTP/1.1
Authorization: Bearer eyJ0eXAiOiJKV1QiLCJhImp0a<truncated>
\end{verbatim}
\end{scriptsize}
\caption{Outgoing network traffic from \texttt{ob.opencampus.mobile} intercepted while performing \sysname. The plaintext email and password for this network request is redacted, and the authorization token is truncated.}
\label{case:study}
\end{figure}

\begin{table}
\begin{minipage}{.45\linewidth}
\centering
\small
\begin{tabular}{l|ccc}
\toprule
& \multicolumn{3}{c}{Consistency Result} \\ \cline{1-4}
Data Flow Types & \textit{Ambiguous} & \textit{Omitted} & \textit{Vague} \\
\midrule
\textbf{Android ID} & 0 & 0 & 1 \\
\textbf{App Name} & 0 & 13 & 0 \\
\textbf{Build Version} & 0 & 12 & 0 \\
\textbf{Cookie} & 0 & 1 & 0 \\
\textbf{Device ID} & 1 & 5 & 6 \\
\textbf{Flags} & 0 & 12 & 0 \\
\textbf{Hardware Information} & 0 & 50 & 1 \\
\textbf{Language} & 0 & 3 & 5 \\
\textbf{Sdk Version} & 0 & 22 & 1 \\
\textbf{Session Information} & 0 & 14 & 0 \\
\textbf{System Version} & 0 & 10 & 0 \\
\textbf{Usage Time} & 0 & 11 & 0 \\
\textbf{User ID} & 1 & 5 & 8 \\
\textbf{VR Field of View} & 0 & 1 & 0 \\
\textbf{VR Movement} & 0 & 8 & 0 \\
\textbf{VR Play Area} & 0 & 2 & 0 \\
\textbf{VR Pupillary Distance} & 0 & 1 & 0 \\
\bottomrule
\textbf{Total} & 2 & 170 & 22 \\
\bottomrule
\end{tabular}
\end{minipage}
\caption{\textit{PoliCheck}\cite{247632} results for the total data flow type for each consistency result using OVRSeen's ontology. The set contains 44 VR apps, with privacy policies, where \sysname was able to trigger outgoing network data flows.}
\label{tab:flowdata_consistency}
\end{table}

\begin{figure}[b]
\scriptsize
\begin{verbatim}
Host: cdp.cloud.unity3d.com
User-Agent: UnityPlayer/2019.2.19f1
Accept-Encoding: deflate, gzip
Accept: */*
Content-Type: application/json
event_count: 1
data_block_id: 64351db8b4d511c2e16a1d97a1907e7b
expired_session_dropped: 0
data_retry_count: 5
continuous_request: 1
request_ts: 1713146923986
X-Unity-Version: 2019.2.19f1
Content-Length: 496

com.aura.Arrows#0.5.5X?????"k??????P??-|{"common":
{"appid":"local.1457b9e91549e2a40a8d759ee2972f52",
"userid":"f213a7998a70e2340aa2f60d6137682c",
"sessionid":5687031950375056550,"platform":"AndroidPlayer",
"platformid":11,"sdk_ver":"u2019.2.19f1","session_count":3,
"localprojectid":"1457b9e91549e2a40a8d759ee2972f52",
"build_guid":"ff76281b9a0c28c46914e0c261de62e1",
"device_id":"b8ba3da0d4458fde63a51772ce84547b"}}
{"type":"analytics.appStart.v1","msg":
{"previous_sessionid":1848577819610040954,
"ts":1713146819046,"t_since_start":11062992}}
com.aura.Arrows#0.5.5Xp??..E.OM:??"oq(?Z?P??yPOST / HTTP/1.1
\end{verbatim}
\caption{Outgoing network traffic from \texttt{com.aura.Arrows} intercepted while performing \sysname.}
\label{case:study2}
\end{figure}

\section{Inclusion of Unreal Engine} \sysname’s internals largely depend on the structure of Unity IL2CPP games, however, the technique of scene loading, event extraction, and execution applies to engines such as Unreal. Unity Scenes are synonymous with Unreal Levels~\cite{epicgamesLevelsUnreal}. UI events are similarly connected using function hooks~\cite{epicgamesDrivingUpdates}. Using scenes/levels to reset the state, identifying UI-events within GameObjects/Blueprints, and extracting the function callbacks are synonymous to both engines. To cover more games within the app stores, Unreal binaries will be considered for future work beyond \sysname.\looseness=-1
\color{black}

\ignore{
\begin{figure}[t]
    \centering
        \centering
        \includegraphics[width=\columnwidth]{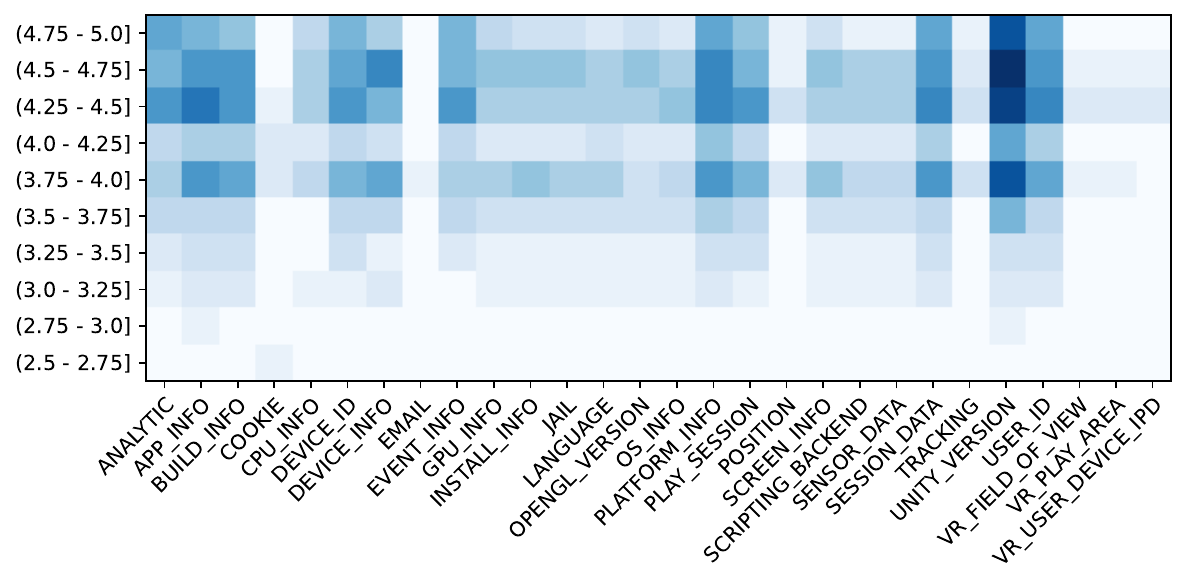}
        \label{fig:free1}
     \vspace{-0.3in}
\caption{A heatmap showing occurrences of sensitive data types per app rating range of both free and paid Meta Quest apps and SideQuest apps. }
\label{fig:sideratingleaks}
\end{figure}
}

\begin{table*}
\scriptsize
\begin{tabular}{c|c|c|c}\toprule
    \multicolumn{1}{c}{\bf API Name} & \multicolumn{1}{|c}{\bf Return Type} & \multicolumn{1}{|c}{\bf Argument Types} & \multicolumn{1}{|c}{\bf Description}  \\
    \midrule
    \hline
    \multicolumn{4}{c}{\bf Classes} \\
    \bottomrule
    \multicolumn{1}{l}{il2cpp\_class\_get\_field\_from\_name} & \multicolumn{1}{l}{field pointer} & \multicolumn{1}{l}{class pointer, string} & \multicolumn{1}{l}{Get class field handle from name.} \\ \hline
    \multicolumn{1}{l}{il2cpp\_class\_get\_fields} & \multicolumn{1}{l}{fields array pointer} & \multicolumn{1}{l}{class pointer, class type} & \multicolumn{1}{l}{Get field handles from class.} \\ \hline
    \multicolumn{1}{l}{il2cpp\_class\_get\_method\_from\_name} & \multicolumn{1}{l}{method pointer} & \multicolumn{1}{l}{class pointer, string, int} & \multicolumn{1}{l}{Get class method handle from method name and parameter count.} \\ \hline
    \multicolumn{1}{l}{il2cpp\_class\_get\_methods} & \multicolumn{1}{l}{methods array pointer} & \multicolumn{1}{l}{class pointer, pointer, class type} & \multicolumn{1}{l}{Get all method handles from class.} \\ \hline
    \multicolumn{1}{l}{il2cpp\_class\_get\_name} & \multicolumn{1}{l}{string} & \multicolumn{1}{l}{class pointer} & \multicolumn{1}{l}{Get name from class handle.} \\ \hline
    \multicolumn{1}{l}{il2cpp\_class\_get\_namespace} & \multicolumn{1}{l}{string} & \multicolumn{1}{l}{class pointer} & \multicolumn{1}{l}{Get class namespace name from class handle.} \\ \hline
    \multicolumn{1}{l}{il2cpp\_class\_get\_parent} & \multicolumn{1}{l}{parent class pointer} & \multicolumn{1}{l}{class pointer} & \multicolumn{1}{l}{Get class' parent handle.} \\ \hline
    \multicolumn{1}{l}{il2cpp\_class\_get\_type} & \multicolumn{1}{l}{class type pointer} & \multicolumn{1}{l}{class pointer} & \multicolumn{1}{l}{Get class type from class handle.} \\ \hline
    \multicolumn{1}{l}{il2cpp\_class\_is\_assignable\_from} & \multicolumn{1}{l}{bool} & \multicolumn{1}{l}{class pointer, other class pointer} & \multicolumn{1}{l}{Test if class is assignable from another class.} \\ \hline
    \bottomrule
    \multicolumn{4}{c}{\bf Fields} \\ 
    \bottomrule
    \multicolumn{1}{l}{il2cpp\_field\_get\_parent} & \multicolumn{1}{l}{parent class pointer} & \multicolumn{1}{l}{class pointer} & \multicolumn{1}{l}{Get parent class of field handle.} \\ \hline
    \multicolumn{1}{l}{il2cpp\_field\_get\_name} & \multicolumn{1}{l}{field pointer} & \multicolumn{1}{l}{string} & \multicolumn{1}{l}{Get field name from field handle.} \\ \hline
    \multicolumn{1}{l}{il2cpp\_field\_get\_offset} & \multicolumn{1}{l}{int32} & \multicolumn{1}{l}{field pointer} & \multicolumn{1}{l}{Get class field offset from field handle.} \\ \hline
    \multicolumn{1}{l}{il2cpp\_field\_static\_get\_value} & \multicolumn{1}{l}{void} & \multicolumn{1}{l}{field pointer, value pointer} & \multicolumn{1}{l}{Get field handle's static value into value pointer.} \\ \hline
    \multicolumn{1}{l}{il2cpp\_field\_get\_type} & \multicolumn{1}{l}{class type pointer} & \multicolumn{1}{l}{field pointer} & \multicolumn{1}{l}{Get type of field from field handle.} \\ \hline
    \multicolumn{1}{l}{il2cpp\_field\_is\_static} & \multicolumn{1}{l}{bool} & \multicolumn{1}{l}{field pointer} & \multicolumn{1}{l}{Test if field is static from field handle.} \\ \hline
    \bottomrule
    \multicolumn{4}{c}{\bf Methods} \\
    \bottomrule
    \multicolumn{1}{l}{il2cpp\_method\_get\_class} & \multicolumn{1}{l}{class pointer} & \multicolumn{1}{l}{method pointer} & \multicolumn{1}{l}{Get residing class handle from method handle.} \\ \hline
    \multicolumn{1}{l}{il2cpp\_method\_get\_name} & \multicolumn{1}{l}{string} & \multicolumn{1}{l}{method pointer} & \multicolumn{1}{l}{Get method name from method handle.} \\ \hline
    \multicolumn{1}{l}{il2cpp\_method\_get\_object} & \multicolumn{1}{l}{object pointer} & \multicolumn{1}{l}{method pointer} & \multicolumn{1}{l}{Get object handle from method handle.} \\ \hline
    \multicolumn{1}{l}{il2cpp\_method\_get\_param\_count} & \multicolumn{1}{l}{uint8} & \multicolumn{1}{l}{method pointer} & \multicolumn{1}{l}{Get number of parameters from method.} \\ \hline
    \multicolumn{1}{l}{il2cpp\_method\_get\_parameter\_name} & \multicolumn{1}{l}{parameters pointer} & \multicolumn{1}{l}{string, uint32} & \multicolumn{1}{l}{Get parameter handle from name and parameter index.} \\ \hline
    \multicolumn{1}{l}{il2cpp\_method\_get\_pointer} & \multicolumn{1}{l}{virtual address pointer} & \multicolumn{1}{l}{method pointer} & \multicolumn{1}{l}{Get virtual address from method handle.} \\ \hline
    \multicolumn{1}{l}{il2cpp\_method\_get\_return\_type} & \multicolumn{1}{l}{class type pointer} & \multicolumn{1}{l}{method pointer} & \multicolumn{1}{l}{Get return type of method.} \\ \hline
    \bottomrule
    \multicolumn{4}{c}{\bf Objects} \\
    \bottomrule
    \multicolumn{1}{l}{il2cpp\_object\_get\_class} & \multicolumn{1}{l}{object pointer} & \multicolumn{1}{l}{class pointer} & \multicolumn{1}{l}{Get class handle from object.} \\ \hline
    \multicolumn{1}{l}{il2cpp\_object\_new} & \multicolumn{1}{l}{object pointer} & \multicolumn{1}{l}{class pointer} & \multicolumn{1}{l}{Create a new object from class handle.} \\ \hline
    \multicolumn{1}{l}{il2cpp\_capture\_memory\_snapshot} & \multicolumn{1}{l}{objects array pointer} & \multicolumn{1}{l}{void} & \multicolumn{1}{l}{Creates a memory snapshot of all GC handles.} \\ \hline
    \bottomrule
\end{tabular}
\caption{IL2CPP Runtime Library API}
\label{tab:il2cppapi}
\end{table*}

\color{black}
\section{Preliminary crash detection}
Because \sysname is agnostic of the privacy/security application, \sysname can be used for additional applications beyond sensitive data exposure, such as crash detection. As such, we separately collected the number of crashes invoked by \sysname and compared it with \textsf{Monkey} as shown in \autoref{tab:crashes}. We notice that a significant number of crashes are \texttt{SEGV\_MAPERR}, which could either be mapping issues between \textit{Frida} and IL2CPP, or critical software crashes. Interestingly, we notice that \textsf{Monkey} was also able to trigger \texttt{SEGV\_MAPERR} crashes, which is unrelated to \textit{Frida} and IL2CPP's mapping, potentially indicating software issues with the game/app itself. We acknowledge that crash detection can be a separate research direction for \sysname, exemplifying \sysname's significance in the VR ecosystem.

\begin{table}[h]
\centering
\small
\begin{tabular}{l|c|c||c|c}
\hline
 & \multicolumn{2}{c||}{Free Apps}  & \multicolumn{2}{c}{Paid Apps} \\ \cline{2-5}
& \multicolumn{4}{c}{\bf{SEGV Code}} \\ \cline{2-5}
  \bf{Tool}  & MAPERR & ACCERR & MAPERR & ACCERR  \\ \hline

Monkey&3&0&0&0 \\  \hline
AutoVR&69&2&5&0 \\ \hline
\end{tabular}
\caption{Aggregated crash count with signal (SIGSEGV) between \textsf{Monkey} and \sysname for both free and paid games.
}
\label{tab:crashes}
\end{table}

\end{document}